# Exact inference under the perfect phylogeny model

Surjyendu Ray [1,*], Bei Jia [3], Sam Safavi [1], Tim van Opijnen [2], Ralph Isberg [4], Jason Rosch [5] and José Bento [1]

[1] Computer Science Department, Boston College, Chestnut Hill, MA 02467, USA
[2] Biology Department, Boston College, Chestnut Hill, MA 02467, USA
[3] Element AI, Toronto, ON M5V 1X2, Canada
[4] Sackler School of Graduate Biomedical Sciences, Tufts University, Boston, MA 02111, USA
[5] Infectious Diseases Department, St. Jude Children's Research Hospital, Memphis, TN 38105, USA

## Abstract

**Motivation:** Many inference tools use the *Perfect Phylogeny Model* (PPM) to learn trees from noisy *variant allele frequency* (VAF) data. Learning in this setting is hard, and existing tools use approximate or heuristic algorithms. An algorithmic improvement is important to help disentangle the limitations of the PPM's assumptions from the limitations in our capacity to learn under it.
**Results:** We make such improvement in the scenario, where the mutations that are relevant for evolution can be clustered into a small number of groups, and the trees to be reconstructed have a small number of nodes. We use a careful combination of algorithms, software, and hardware, to develop EXACT: a tool that can explore the space of *all* possible phylogenetic trees, and performs *exact inference* under the PPM with noisy data. EXACT allows users to obtain not just the most-likely tree for some input data, but exact statistics about the distribution of trees that might explain the data. We show that EXACT outperforms several existing tools for this same task.
**Availability:** https://github.com/surjray-repos/EXACT
**Contact:** raysc@bc.edu

## 1 Introduction

Phylogeny reconstruction from different forms of mutation data is important for many evolutionary-related studies. For example, in antimicrobial resistance research, one often seeks the evolutionary history of bacteria as they develop resistance to one, or a combination of, antibiotics (Palmer and Kishony, 2013). In cancer research, one often wants to determine the proportions of clonal sub-populations in tumor samples, from heterogeneous mixtures of such sub-populations, where, as cancer progresses, intra-tumor heterogeneity arises through evolution (Merlo *et al.*, 2010; Ruiz *et al.*, 2011; Kreso *et al.*, 2013). Many phylogeny inference tools have been proposed for varied tree inference tasks. These tools can be categorized in terms of the underlying models assumed, the input data types, the size of the problems that can be tackled, and the mathematical principles used for reconstruction.

A canonical evolutionary model that underlies several tools is the Perfect Phylogeny Model (PPM) (Hudson, 1983; Kimura, 1969). It assumes that it is (almost) impossible for the same position in the genome to mutate twice, and hence that, mutations (almost) always only accumulate. Proposed extensions to the PPM have been made (Bonizzoni *et al.*, 2014; Hajirasouliha and Raphael, 2014), and tools that go beyond the PPM model also exist. For example, some tools such as MEDICC (Schwarz *et al.*, 2014), TuMult (Letouzé *et al.*, 2010), and FISHtrees (Gertz *et al.*, 2016) relax the PPM model to allow the number of copies of a given genetic element, or ploidy, to both increase, or decrease, thus effectively allowing the removal of mutations. However, tools based on the PPM alone are still very important, and an active area of research, as can been seen from Table 8 in Sec. 6 in the Supp. file. Among these tools, the input type varies. Sequencing data can come from a single sample, or multiple samples, each sample containing changes in variant allele frequencies (VAFs) of single-nucleotide mutations/variants (SNVs); i.e., the fraction of bacterial or tumor cells that contain each mutation. Other methods use single cell sequencing data as input.

In this paper, we develop a tool to learn phylogenetic trees from multiple samples of VAFs under the PPM. Table 8 in Sec. 6 in the Supp. file includes all of the algorithms that are of the same type as our method and that are (i) state-of-the-art, and (ii) actively maintained and used, namely, PhyloSub (Jiao *et al.*, 2014), AncesTree (El-Kebir *et al.*, 2015b), CITUP (Malikic *et al.*, 2015), PhyloWGS (Deshwar *et al.*, 2015), Canopy (Jiang *et al.*, 2016), SPRUCE (El-Kebir *et al.*, 2016), rec-BTP (Hajirasouliha *et al.*, 2014), PASTRI (Satas and Raphael, 2017) and LICHeE (Popic *et al.*, 2015). A good review of all of these methods can be found in Schwartz and Schäffer (2017) and Holder and Lewis (2003).

Although the PPM is simple, inferring the best PPM-based model that explains a data set with noisy, or incomplete VAFs is hard (El-Kebir *et al.*, 2015b). This algorithmic hardness has influenced the development of tools, introducing on them inexactness that, at least in some cases, could be avoided. See Section 2, for details on different algorithms.

Typical hardness results do not exclude the possibility of solving particular instances of a problem, e.g., for small-sized instances. At the same time, in many relevant settings, and in particular in several of the experiments in the papers cited above (See Table 8 in Sec. 6 in Supp. file), the trees inferred have a relatively small number of resultant nodes. This begs the question: why are there no tools that exactly solve the PPM under a noisy setting for small but practical, biologically relevant problem sizes?

We use a careful combination of algorithm design, software development, and choice of computing hardware, to provide such a tool. We name it EXACT. In a nutshell, we (i) use a novel algorithm that exactly evaluates the data-fitness cost of small trees in a fraction of a millisecond, (ii) implement this naturally-recursive algorithm in C, in a non recursive form and using only low-level primitives to achieve greater speed, and (iii) embed this code into a CUDA program that uses Graphical Processing Units (GPUs) to explore the full space of possible trees. E.g. we can explore billions of trees in a couple of hours using a single GPU. Our tool has the following unique advantages:
**1.** When the input data is noisy, it is not enough to return one, or a few, trees that approximately maximize the posterior probability, but rather, we want to return exact likelihood scores for many trees, so that the user can perform statistical queries with confidence. Unlike existing tools, EXACT can do this, since it scans the complete space of trees, and, for each tree, exactly computes its likelihood;
**2.** The tool needs to be practical and return an answer within a reasonable amount of time using a reasonable amount of a possibly diverse set of computation resources. EXACT can easily exploit multiple CPUs, multiple GPUs, or multiple machines, and even using a single consumer-GPU, solve problems of practical size fast, which no other tool can do.
**3.** Since we are doing exact inference under the PPM, our tool serves as a benchmark for existing, and future tools. Useful tools, including the ones that scale to large problem sizes, or use more complex evolution models, should be able to infer close-to-optimal phylogeny trees under the canonical PPM for small but relevant problem sizes, which we can check with EXACT.

## 2 Background and Related Work

We review the main technical limitations of the algorithms that solve the problem of inferring phylogenetic trees under the PPM model with noisy VAFs as input. AncesTree and CITUP require solving an Integer/Quadratic Linear Program (ILP/IQP), which is an NP-hard problem in general, and which is approximated via branch-and-bound heuristics. By nature, they return only "best-fitting" solutions, and do not give information about the likelihood of other solutions. BitPhylogeny, Canopy, and PhyloWGS are sampling-based methods, which in turn sample distributions using MCMC strategies; LICHEE is also another sampling-based methodology. They can, in principle, return information about the likelihood of different solutions, although among the three, PhyloWGS is the only one that does it out of the box. However, they are by nature, inexact, and in practice, even for small problems and under noisy data, they need to run for a long time to achieve some reasonable accuracy (cf. PhyloWGS run-time in Section 4.1). PASTRI is also a sampling-based method. It is not based on MCMC methods but on importance sampling. The authors claim it is more accurate than other MCMC-based methods, but do not report in their paper any directly comparison with these methods. AncesTree, SPRUCE and Lichee reduce the space of possible trees by building directed graph structures that encode which mutations can precede other mutations. However, these structures can only exclude sub-optimal trees with certainty when the noise in the data is bounded (with known bound), which is not always true. Furthermore, even in this setting, sometimes these tools need to take heuristic decisions on whether a given ancestral relation should be excluded or not. TuMult, rec-BTP and MEDICC use greedy algorithms, and find sub-optimal solutions.

## 3 Materials and Methods

The PPM assumes that the same position in the genome never mutates twice, hence mutations only accumulate, and are never lost. Consider a population of organisms evolving under the PPM. The evolution process can be described by a labeled rooted tree, $T = (r, \mathcal{V}, \mathcal{E})$, where $r$ is the root, i.e., the common oldest ancestor, the nodes $\mathcal{V}$ are the mutants, and the edges $\mathcal{E}$ are mutations acquired between older and younger mutants. Since each position in the genome only mutates once, we can associate with each node $v \neq r$, a unique mutated position, the mutation associated to the ancestral edge of $v$. By convention, let us associate with the root $r$, a null mutation that is shared by all mutants in $T$. This allows us to refer to each node $v \in \mathcal{V}$ as both a mutation in a position in the genome (the mutation associated to the ancestral edge of $v$), and a mutant (the mutant with the fewest mutations, that has a mutation $v$). Hence, without loss of generality, $\mathcal{V} = \{1, \ldots, q\} \triangleq [q]$, $\mathcal{E} = \{2, \ldots, q\} = 1 + [q-1]$, where $q$ is the length of the genome, and $r = 1$ refers to both the oldest common ancestor and the null mutation shared by all.

Consider the problem of inferring how mutants of a common ancestor evolve. In this paper our input data is the frequency with which different positions in the genome mutate across multiple samples. Consider a sample $s$, one of $p$ samples, obtained at a given stage of the evolution process of an organism. This sample has many mutants, some with the same genome, some with different genomes. Let $F \in \mathbb{R}^{q \times p}$ be such that $F_{v,s}$ is the fraction of genomes in $s$ with a mutation in position $v$ in the genome. Let $M \in \mathbb{R}^{q \times p}$ be such that $M_{v,s}$ is the fraction of mutant $v$ in $s$. By definition, the columns of $M$ must sum to 1. Let $U \in \{0,1\}^{q \times q}$ be such that $U_{v,v'} = 1$, if and only if mutant $v$ is an ancestor of mutant $v'$, or if $v = v'$. We denote the set of all possible $U$ matrices, $M$ matrices and labeled rooted trees $T$, by $\mathcal{U}, \mathcal{M}$ and $\mathcal{T}$, respectively. Any tree $T \in \mathcal{T}$ can be represented through a binary matrix $T$, where $T_{i,j} = 1$ if and only if node $i$ is the parent of node $j$. Henceforth, we let $\mathcal{T}$ denote the set of all such binary matrices $T$. Note that $U$ uniquely determines $T$, and vice versa.

Lemma 3.1. *Consider an evolutionary tree and its corresponding matrices $T \in \mathcal{T}$ and $U \in \mathcal{U}$, we have that $U = (I - T)^{-1}$.*

The proof of Lemma 3.1 is standard, and we include its proof in Supp. Mat. 12. See Fig. 2 for an illustration of the PPM model. The PPM implies

$$F = UM. \tag{1}$$

Our goal is to infer clonal evolution, i. e., $M$ and $U$, from mutation-frequencies, i.e., $F$. When $F$ is observed cleanly, it is easy to find solutions to this problem. As explained in El-Kebir *et al.* (2015b), we can build a Directed Acyclic Graph (DAG), where mutant type $i$ connects to mutant type $j$ iff $j$ cannot be an ancestor of $i$, which happens iff $F_{j,s} < F_{i,s}$ for some sample $s$. Any spanning tree of this DAG is a valid evolution tree $T$. From $T$, we can build $U$, and then read $M = U^{-1}F$. This procedure can be completed in $\mathcal{O}(q^2 p)$ steps. Variations of this procedure are the basis of AncesTree, SPRUCE, and LICHeE.

### 3.1 Inference

In practice, since sequencing can be a noisy process, we only have access to a noisy version of a few components of $F$, which we denote by $\hat{F}_O$, where $O$ is the set of observed pairs of genome positions and samples. Our goal is to recover $U$ and $M$ from $\hat{F}_O$. We explain our inference procedure from a Bayesian perspective, although other justifications are possible.

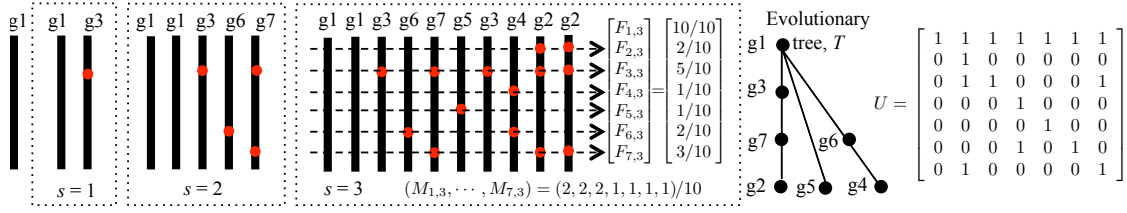

**Fig. 1.** Black lines are genomes, and red dots mutations. $gi$ is the type of mutant with fewest mutations with position $i$ mutated. The mutation in the null position $i=1$, is shared by all mutants, and $g1$ is the organism's genome before evolution starts. In sample $s=3$, $2/10$ of the mutants are of type $g2$, hence $M_{2,3}=2/10$, and $3/10$ of the mutations occur in position 7, hence $F_{7,3}=3/10$. The tree shows the mutants' evolution. Each tree node represents both a mutation in position $i$ and a mutant type $gi$. Matrix $U$ has a one in (row,column)=(3, 2) because $g3$ is an ancestor of $g2$. The PPM implies that $F_{\cdot,3}=UM_{\cdot,3}$, where $F_{\cdot,3}$ and $M_{\cdot,3}$ are the vectors displayed above in sample $s=3$.

Let $f_{\hat{F}_O,F,U,M}$ be the joint probability density function of the random variables $\hat{F}_O, F, U, M$, which satisfies

$$f_{\hat{F}_O|F,U,M} = \prod_{(i,s)\in O} f_{\hat{F}_{i,s}|F_{i,s}}, \quad (2) \qquad F = UM, \quad (3)$$

$$\hat{F}_{i,s} - F_{i,s} \sim \mathcal{N}(0, \Sigma_{i,s}), \quad (4) \qquad f_{M,U} = Cf_U, \quad (5)$$

for some constant $C$, and an arbitrary known set of variances $\{\Sigma_{i,s}\}$, which we assume are all strictly positive. In other words: the PPM holds; the measurement of each observed $F_{i,s}$ is independent of the other measurements; the measurement noise is zero mean Gaussian with variance $\Sigma_{i,s}$; and there is no prior knowledge on the distribution of mutants. Under these assumptions, the max. likelihood estimation of $U, M$ given $\hat{F}_O$ is

$$\max_{U\in\mathcal{U}} \mathcal{I}(\mathcal{C}(U;\hat{F}_O)) + \mathcal{Q}(U), \quad (6)$$

where $\mathcal{Q}(U)$ models the prior knowledge on the possible trees, $\mathcal{C}(U;\hat{F}_O)$ models the data fitting cost, and, for now, $\mathcal{I}(x)=-x$. With a slight abuse of notation, we will sometimes write $\mathcal{C}(U;\hat{F}_O)$ as $\mathcal{C}(T;\hat{F}_O)$. Given the assumptions (2)-(5), $\mathcal{C}(U;\hat{F}_O)$ can be expressed as

$$\mathcal{C}(U;\hat{F}_O) = \min_{M, F\in\mathbb{R}^{q\times p}} \|D\otimes (\hat{F}_O - F_O)\|^2 \quad (7)$$

subject to $F=UM, M\geq 0, M^\top \mathbf{1} = \mathbf{1}$,

where $D_{i,s} = \Sigma_{i,s}^{-1}$, $\otimes$ denotes element wise product, $\|\cdot\|$ is the Frobenius matrix norm, $\top$ denotes transpose, and $\mathbf{1}$ is the vector of all ones.

Our tool solves (6) for generic (fast-to-compute) functions $\mathcal{I}$ and $\mathcal{Q}$ via full enumeration of all possible $U$ matrices. This endows it with the unique properties described in Section 1. One key step to do so is a new, very fast algorithm to compute the cost $\mathcal{C}(U;\hat{F}_O)$ for each matrix $U$. Note that (7) decouples into $p$ independent problems, one per sample $s$. Therefore, **from now on, we assume that we only have one sample, i.e., $p=1$**. This also allows us to slightly simplify notation, and write,

$$\mathcal{C}(U;\hat{F}_O) = \min_{M, F\in\mathbb{R}^q} \|D(\hat{F}_O - F_O)\|^2 \quad (8)$$

subject to $F=UM, M\geq 0, M^\top \mathbf{1} = 1$,

where $F$ and $M$ are now vectors, the norm is the Euclidean norm, and $D$ is now a strictly positive diagonal matrix with $\{\Sigma_i^{-1}\}_{i\in O}$ in the diagonal.

### 3.2 Missing observations

Problem (8) can be further simplified to only include variables related to observed mutated positions. Without loss of generality let $O = [|O|]$.

**Theorem 3.2.** *Given $O \subseteq [q]$, let $H = [q]\setminus O$. Consider the following block representation of $M$ and $U$: $M = [M_O; M_H]$, and $U = [U_{O,O}, U_{O,H}; U_{H,O}, U_{H,H}]$. Let $\tilde{M}_O = M_O + (U_{O,O})^{-1}U_{O,H}M_H$. We have that*

$$\mathcal{C}(U;\hat{F}_O) = \min_{\tilde{M}_O, F_O \in \mathbb{R}^{|O|}} \|D(\hat{F}_O - F_O)\|^2 \quad (9)$$

*subject to* $F_O = U_{O,O}\tilde{M}_O, \tilde{M}_O \geq 0, \tilde{M}_O^\top \mathbf{1} \stackrel{=}{\leq} 1, \quad (10)$

*where in the last constraint, we choose $=$ if $O$ includes the root of the tree associated with $U$, and $\leq$ otherwise.*

### 3.3 Very fast algorithm to compute the cost of each tree

We describe our fast algorithm to compute $\mathcal{C}(U;\hat{F}_O)$ for "=" in the constraint (10). It is a simple modification to deal with the case where we use "≤" in (10), which we explain in Supp. Mat. Sec. 9. The algorithm is an important extension of a simpler algorithm that we introduced at the 2018 conference for the Advances in Neural Information Processing Systems (Jia *et al.*, 2018). In this section, for simplicity, we drop the $O$ subscript in all the variables, and also drop the $\tilde{}$ so that, e.g., $\tilde{M}_O$ here reads $M$.

The starting point is a dual formulation for problem (9). Let $\bar{i}$ be the closest ancestor of $i$ in $T=(r, \mathcal{V}, \mathcal{E})$, the phylogenetic tree rooted at $r$ associated with $U$. Let $\Delta i$ be the set of all the ancestors of $i$ in $T$, plus $i$. Let $\partial i$ be the set of the children of $i$ in $T$.

**Theorem 3.3.** *Problem* (9) *can be solved by solving*

$$\min_{t\in\mathbb{R}} t + \mathcal{L}(t), \quad (11)$$

$$\mathcal{L}(t) = \min_{Z\in\mathbb{R}^q} \frac{1}{2}\sum_{i\in\mathcal{V}} D_{ii}^{-2}(Z_i - Z_{\bar{i}})^2 \quad (12)$$

*subject to* $Z_i \leq t - N_i, \forall i \in \mathcal{V}, \quad (13)$

*where $N_i = \sum_{j\in\Delta i} D_{jj}^2 \hat{F}_j$, and, by convention, $Z_{\bar{i}} = 0$ for $i = r$. In particular, if $t^*$ minimizes* (11), $Z^*$ *minimizes* (12) *for $t = t^*$, and $M^*, F^*$ minimize (9), then $M_i^* = D_{ii}^{-2}(-Z_i^* + Z_{\bar{i}}^*) + \sum_{r\in\partial i} D_{rr}^{-2}(Z_r^* - Z_{\bar{r}}^*)$, and $F_i^* = D_{ii}^{-1}(-Z_i^* + Z_{\bar{i}}^*), \forall i \in \mathcal{V}$. Also, $t^*, M^*, F^*, Z^*$ are unique.*

A formal derivation of Theorem 3.3 is in Supp. Mat. 12. Starting from Theorem 3.3, we now explain in words the main foundation behind our algorithm. We compute $Z^*(t)$ for each $t$, i.e., the minimizer of problem (11). At the same time, from $Z^*(t)$, we compute $\mathcal{L}(t)$, and by taking a derivative, we also compute $\mathcal{L}'(t)$ for each $t$. Using $\mathcal{L}'(t)$ we find $t^*$ such that $1 + \mathcal{L}'(t^*) = 0$, which is the solution to (11). Finally, once $t^*$ is found, we compute $Z^*(t^*)$ and use the relations in Theorem 3.3 to obtain $M^*$ and $F^*$, the minimizers of (9). Several of the quantities that we deal with are piecewise smooth (cf. Lemma 10.4 in App. 10). We assume that all derivatives are taken from the left, i.e., $f'(t) \triangleq \lim_{s\uparrow t}(f(t) - f(s))/(t-s)$.

The following gives more detail on how we compute $Z^*(t)$, and the other aforementioned quantities. The minimizer and minimum of problem (12), i.e., $Z^*(t)$ and $\mathcal{L}(t)$, change continuously with $t$ (cf. Lemmas 10.1 and 10.2 in App. 10). Furthermore, apart from a few special values of $t$, the nodes $i \in \mathcal{V}$ for which the constraints (13) are active, and which we denote by $\mathcal{B}(t)$, do not change (cf. Lemma 10.5 in App. 10). In each interval of $t$ for which the set of active constraints do not change, problem (12) is a convex quadratic problem, where some of the variables (the ones corresponding to the active constraints) change linearly with $t$. Hence, $Z^*(t)$ is a piece-wise linear function of $t$ (cf. Lemma 10.4 in App. 10). As $t$ decreases, there is a change in linear regime every time a new constraint becomes active, i.e., every time $Z^*(t)$ intersects $t - N_i$. Since the rate of change of $Z^*(t)$ is controlled by the factor multiplying $t$ in (13), and since this factor is 1, the rate of change of each component $Z^*(t)_i$ is, at most, 1 (cf. Lemma 11.2 in App. 10). Hence, as $t$ decreases, and once $Z_i^*(t) = t - N_i$ for some $i$, the equality constraint will hold for all other smaller values of $t$. In turn, this implies that there can be at most $q$ changes in linear regime (cf. Lemmas 10.5 and 10.6 in App. 10).

Therefore, to fully describe $Z^*(t)$, we only need to compute $Z^*(t)$ at the different *critical values* $t_1 > t_2 > \cdots > t_k$, $k \leq q$, at which the piece-wise linear continuous function $Z^*(t)$ changes linear regime. We do so sequentially (**line 3 in Alg. 1**). For some large $t_0$, we can easily compute $Z^*(t_0) = 0$ and $Z'^*(t_0) = 0$. We can then extrapolate $Z^*(t)$ for smaller values of $t$ until we find the largest $t_1 < t_0$ for which $Z_i^*(t_1) = t_1 - N_i$ for some nodes $i$. We store all nodes $i$ whose constraint (13) is active at $t_1$ in a set $\mathcal{B}(t_1)$ (**line 2 in Alg. 1**). At this point, we compute a new slope $Z'^*(t_1)$ (Procedure *ComputeRates* in **line 4 in Alg. 1**) (we explain in App. 11 how we can do this using only $\mathcal{B}(t_1)$), and, using this slope, we extrapolate $Z^*(t_1)$ for smaller values of $t$ until we find the largest $t_2 < t_1$ for which $Z_i^*(t_1) = t_1 - N_i$ for some nodes $i$ (**lines 5-6 in Alg. 1**). We store all nodes $i$, whose constraint (13) is active at $t_2$ in a set $\mathcal{B}(t_2)$ (**line 7 in Alg. 1**). After we compute the new critical point, we also evaluate $Z^*(t_2)$ via extrapolation (**line 8 in Alg. 1**). We keep repeating this process.

As we compute $Z^*(t_i)$ and $Z'^*(t_i)$ for different critical points $t_i$, we also compute $\mathcal{L}'(t_i)$ (**line 9 in Alg. 1**), and $\mathcal{L}''(t_i)$ (**line 4 in Alg. 1**). We stop this process once we find the first critical point $t_{i+1}$, for which $\mathcal{L}'(t_{i+1}) < -1$ (**line 10 in Alg. 1**). Since $\mathcal{L}'(t)$ is also piece-wise linear and continuous (the derivative of a piece-wise quadratic continuous function), we can interpolate $\mathcal{L}'$ and $Z^*$ between $t_{i+1}$ and $t_i$ to find $t^*$ such that $\mathcal{L}'(t^*) = 0$, and then $Z^*(t^*)$ (**lines 12-13 in Alg. 1**). Once $Z^*(t^*)$ is found, $M^*$ and $F^*$ are computed using the relations in Theorem 3.3 (**line 14 in Alg. 1**). The algorithm is described fully in Algorithm 1.

---

**Algorithm 1** Compute tree cost (input: $T$ and $\hat{F}$; output: $M^*$ and $F^*$)

1: $N_i = \sum_{j \in \Delta_i} D_{jj}^2 \hat{F}_j$, for all $i \in \mathcal{V}$
2: $i=1, t_i = \max_r \{N_r\}, \mathcal{B}(t_i) = \arg\max_r \{N_r\}, Z^*(t_i) = \mathbf{0}, \mathcal{L}'(t_i) = 0$.
3: **for** $i = 1$ to $q$ **do**
4: $\quad (Z'^*(t_i), \mathcal{L}''(t_i)) = \text{ComputeRates}(\mathcal{B}(t_i), T)$
5: $\quad P = \{P_r : P_r = \frac{N_r + Z_r^*(t_i) - t_i Z_r'^*(t_i)}{1 - Z_r'^*(t_i)}$ if $r \notin \mathcal{B}(t_i) \wedge (t_r < t_i)$; and $P_r = -\infty$ o.w.
6: $\quad t_{i+1} = \max_r P_r$
7: $\quad \mathcal{B}(t_{i+1}) = \mathcal{B}(t_i) \cup \arg\max_r P_r$
8: $\quad Z^*(t_{i+1}) = Z^*(t_i) + (t_{i+1} - t_i) Z'^*(t_i)$
9: $\quad \mathcal{L}'(t_{i+1}) = \mathcal{L}'(t_i) + (t_{i+1} - t_i) \mathcal{L}''(t_i)$
10: $\quad$ **if** $\mathcal{L}'(t_{i+1}) < -1$ **then break**
11: **end for**
12: $t^* = t_i - \frac{1 + \mathcal{L}'(t_i)}{\mathcal{L}''(t_i)} \quad \triangleright$ If "$\leq$" in (10), then also do $t^* = \max\{t^*, 0\}$
13: $Z^* = Z^*(t_i) + (t^* - t_i) Z'^*(t_i)$
14: **return** $M^*, F^*$

---

**Theorem 3.4.** *Algorithm 1 outputs the solution to* (9) *in $\mathcal{O}(q^2)$ steps, and requires $\mathcal{O}(q)$ memory.*

Algorithm 1 is not an iterative algorithm that gradually approximates $\mathcal{C}(U; \hat{F}_O)$, it always computes $\mathcal{C}(U; \hat{F}_O)$ with arbitrary precision after $\mathcal{O}(q^2)$ steps, the only limiting factor being the machine precision used when coding it (32bits, 64bits, 128bits, etc.). No current tool does this. Also, it is immediate to see from Alg. 1 that if $\hat{F}_0$ are fractional numbers, then all of the variables in Alg. 1, and the final solution, will be fractional numbers. In this case, we achieve infinite precision in a finite number of steps.

### 3.4 Exploring all trees

We have seen how to compute $\mathcal{C}$ very fast. If $\mathcal{Q}$ and $\mathcal{I}$ are also fast to compute, then we can solve (6) by comparing the cost of all possible $U$ matrices, i.e., all trees. This is unlike many other algorithms, which use heuristics to avoid searching the space of all possible trees (See Table 8 in Sec. 6 in Supp. file). An exhaustive search is necessary when the noise in $\hat{F}_O$ is high, or when we want to collect accurate statistics about the likelihood of different trees. E.g. we might want to list the top 100 most likely trees given $\hat{F}_O$, which no current software can guarantee to do without some form of approximation.

We consider two scenarios: **A)** we scan all of the rooted trees with $q$ nodes; **B)** we scan all of the spanning trees of a user-input rooted DAG with $q$ nodes. Our tool is agnostic to how the DAG was built. The user should have in mind, however, that building this DAG as in AncesTree or LiCHeE can result in inaccuracies, as explained in Section 2. Without loss of generality, we can assume that we know the root of the tree to be learnt, since we can introduce a null mutation $r$, common to all mutants, that has a frequency of occurrence $\hat{F}_r = 1$, and that will be the root of $T$.

In both cases, the main idea behind our search is as follows:
**1.** Create a set of consecutive indices $\mathcal{G} = [|\mathcal{G}|]$, where $|\mathcal{G}|$ is the number of trees that need to be scanned; **2.** Divide $\mathcal{G}$ into $h$ subgroups, $\{\mathcal{G}_i\}_{i=1}^h$, not necessarily of the same size; **3.** Assign each $\mathcal{G}_i$ to a computing unit, i.e., the CPUs or a GPU of a specific machine; **4.** Each computing unit assigns the indices in $\mathcal{G}_i$ to its different cores, and, in parallel, **5.** Each core sequentially **transforms each index assigned to it into a tree** $T$, computes $\mathcal{C}(U; \hat{F}_O)$, and returns the $k$ trees with the lowest cost $\mathcal{C}$ among the trees tested (using binary heap for this purpose); **6.** The $k$ trees returned by different cores in each computing unit are merged, and each computing unit returns the $k$ trees with the lowest cost among all trees tested by that unit; **7.** Finally, a master script merges the different sets of $k$ trees computed by different computing units, and returns $k$ trees with the lowest cost across all possible trees.

What is left to explain is the line in bold in step **5**. In scenario **A**, for $q$ nodes, there are $|\mathcal{T}| = q^{q-2}$ trees rooted in $r$ (w.l.o.g. $r$ is assumed to be known) (Cayley, 1889), and we map indices into trees using *Prüfer sequences*. A Prüfer sequence on $q$ labels, is a sequence of length $q - 2$, where each number in the sequence takes values in $[q]$. We first convert each of the $|\mathcal{T}|$ available indices into a Prüfer sequence. If $x$ is an index, and $y$ its Prüfer sequence, we obtain $y = (y_1, \ldots, y_{q-2})$ from $x$ by setting $k = 1$ and repetitively doing $y_k \leftarrow 1 + (x \mod q); x \leftarrow \text{floor}(x/q); k \leftarrow k + 1$. We then use Prüfer's algorithm (Prufer, 1918) to get a tree from $y$ in $\mathcal{O}(q^2)$ steps. Although there are $\mathcal{O}(q)$ methods (Micikevicius *et al.*, 2006), we use Prüfer's since this step is not the bottleneck. In scenario **B**, if $\{c_i\}_{i \in [q] \setminus r}$ is the list of the number of parents of each node in the DAG (except the root), then there are $|\mathcal{T}| = \prod_{i \in [q] \setminus r} c_i$ spanning trees. Basically, we need to choose one parent for each node $i \neq r$. Assuming w.l.o.g. that $r = q$, for each index $x \in [|\mathcal{T}|]$, we can get a list of parents $\{y_i\}$ that each node $i$ should connect to, and hence a spanning tree, by setting $k = 1$ and repetitively doing $y_k \leftarrow 1 + (x \mod c_k); x \leftarrow \text{floor}(x/c_k); k \leftarrow k+1$.

### 3.5 Extensions

#### 3.5.1 Clustering

Just like in many other tools, EXACT also allows clustering different mutated positions based on the values of $\hat{F}_O$. Assume for simplicity that $O = [q] \times [p]$. Given the vectors $\{\hat{F}_{v,\cdot} \in \mathbb{R}^p, \forall v \in [q]\}$, we partition $[q]$ into $\ell$ sets $S_1, \ldots, S_\ell$, such that $\hat{F}_{v,\cdot}, v \in S_i$, and $\hat{F}_{v',\cdot}, v' \in S_j$, are similar if $i = j$, and dissimilar if $i \neq j$. Currently, EXACT does this using *kmeans* clustering (Lloyd, 1982), and ignoring the uncertainty vector $\text{diag}(D)$, just like in CITUP, but other clustering techniques can be used. The clustering is done as a pre-computation step on the observed values $\hat{F}_O$. However, once the de-noised values $F$ are computed, we could, in principle, re-cluster the positions $[q]$ using $F$, and then re-compute $F$, etc. Once the clusters are computed, it is as if we have a new number of mutated positions, and we use the centers of each cluster as the new input vector of frequencies to our algorithm. Doing this, and using an appropriate uncertainty matrix $D$ at each step, will avoid some of the shortcomings of directly clustering with VAFs that are discussed in Section 3.2 in El-Kebir *et al.* (2015b).

With a slight abuse of notation, in the rest of this section, we denote the new number of clusters by $q$ and the new input vector by $\hat{F}_O$. We choose the value of $q$ using a Bayesian information (BIC) criterion (Schwarz, 1978).

Let $\mathcal{H}(\hat{F}_O)$ be the value of (6), we find $q$ by solving

$$\min_q 2\mathcal{H}(\hat{F}_O) + p(q-1)\log q^*, \quad (14)$$

where $q^*$ is the number of mutated positions before clustering is performed. This is similar to what is done in CITUP (cf. eq. 6 in Malikic *et al.* (2015)).

In our numerical experiments, we observe that using this pre-clustering and BIC criterion leads to good performance (cf. Section 4).

### 3.5.2 Other objective functions
Although the objective (7) looks simple, it allows us to extend our tool to other objectives, as we describe in Section 7 in the Supp. file.

## 4 Numerical Results

We now test how well different algorithms infer phylogenies from data.
**Algorithms:** The algorithms tested are EXACT (our algorithm), as well as Canopy (McGranahan and Swanton, 2017), AncesTree (El-Kebir *et al.*, 2015b), PhyloWGS (Deshwar *et al.*, 2015), and CITUP (Malikic *et al.*, 2015). We use these algorithms because they (i) come with publicly available code; (ii) accept as input the frequency of mutations per genome-position for different samples, and output phylogenetic trees; (iii) perform inference under PPM-type assumptions; and (iv) cover the main techniques used in the literature, namely, probabilistic inference via sampling methods, combinatorial optimization via ILP/IQP, and exhaustive searches.
**Data sets:** We use the data sets provided by El-Kebir *et al.* (2015b), which contain 36 real data files, and 90 simulated pairs of input/ground truth files We use these data sets because they have been previously used to compare other algorithms. Each file encodes a different $\hat{F}_O$ via sequencing counts. AncesTree, Canopy and PhyloWGS directly take as input sequencing counts, while CITUP and EXACT take as input $\hat{F}_O$. If $X$ and $Y$ are the counts in position $v$, sample $s$, for the mutated and reference allele respectively, then the following will hold $\hat{F}_{v,s} = \frac{2X}{X+Y} \in [0,1]$. Some real data set input files contain empty cells in the count matrix. These cells are treated as 0.
**Code:** EXACT was run on a NVIDIA Quadro P5000 GPU. All other code was run on a computer with Intel Xeon E5-2660 v4 CPUs at 2.0GHz. CITUP used 32 cores. Other tools used only a single CPU core. Code for these tools can be found through links in each respective paper. Only our tool can process all of the data sets out-of-the-box. Table 1 lists the data set files on which different tools failed (cf. App. Section 13 for details). We had to run CITUP and Canopy multiple times on some input files to get a valid output. The performance of each tool is evaluated on the files, on which it runs to completion. The comparison of the output of any pair of tools is done on the files on which both tools run to completion. Except for CITUP and EXACT, the different tools do not always output the same result when run repeatedly on the same input. We use the default parameters in each tool, except in CITUP, where we increase the maximum size of the tree to fit, from 5 to 9. For EXACT, for each input file, we test all of the trees of sizes 7 to 10 for synthetic data, and of sizes 6 to 9 for real data, and in (6) we set $D = 1/0.06, \mathcal{I}(x) = x$, and $\mathcal{Q} = 0$.

| Data\Tool | AncesTree | Canopy | CITUP | PhyloWGS |
|---|---|---|---|---|
| Synthetic | | 66,68,70,83,85 | | 35 |
| Real | 1,2,9-13,15,17-20 | 1-22,30,31 | 7,8 | |

Table 1. File IDs on which tools failed. For real data, we compute the ID by listing all the file names in increasing lexicographical order; for synthetic data, we compute the ID by visiting all the synthetic data folders in increasing lexicographical order, and inside each folder, visiting all file names in increasing lexicographical order.

Three aspects affect algorithms' performances: (i) which mathematical models better explain the data; (ii) which formulation is a given algorithm trying to perform inference on; and (iii) how close each algorithm is to performing exact inference on the assumed formulation. Regarding (i), the real data does not follow the PPM. The synthetic data follows a noisy PPM at the level of the clustered VAFs, which are not directly accessible. The actual input data consists of non-clustered mutations and does not follow a noisy PPM. Regarding (ii), all tools assume some variation of the infinite sites assumption, the basis of the PPM. The tools that extend the canonical PPM do so to deal with copy number aberrations (CNA), e.g. in PhyloWGS and Canopy. However, in all of the comparisons that we make, we only test the different algorithms in a scenario where there are no CNAs present. The idea is to test how well the different algorithms perform in a situation where the data is as close as possible to the canonical PPM. Regarding (iii), AncesTree and CITUP try to perform exact inference on the PPM with a noisy input, but in the end require solving an ILP, or a QIP, which, in turn, is solved heuristically. Canopy and PhyloWGS perform approximate inference using sampling methods to sample from the posterior of the tree distribution given the noisy observed data. EXACT does exact inference from noisy observations by completely enumerating all possible trees and, for each tree, exactly computing a likelihood cost.

### 4.1 Results on synthetic data

Each input file encodes the (noisy) mutation counts and reference allele counts in 100 different positions of 4, 5 or 6 samples. The counts simulate a sequencing coverage of 50, 100, and 1000 times. There are 10 input files per sequencing coverage and sample size. Each ground truth file contains a grouping of the 100 positions into 10 sets (clusters), and a phylogenetic tree on these 10 clusters. Appendix C of El-Kebir *et al.* (2015a) has additional details on the creation of the synthetic data.

We test the performance of each tool by comparing its output tree (the tree which the tool deems to best explain the data) with the ground truth tree, using four types of errors. Tree nodes can be associated to multiple (clustered) genome positions, to a single position, or to none. To explain the four error types, we introduce the following notation. Let $i$ and $j \neq i$ be any pair of mutated positions. We define $Cl_g(i,j)$ (resp. $Cl_o(i,j)$) to be 1 if $i$ and $j$ belong to the same cluster in the ground truth (resp. output) tree, or 0 if they both belong to some but different clusters. If either $i$ or $j$ is not in any cluster in the output tree, then we set $Cl_o(i,j)$ to Nan. When comparing $Cl_o$ to $Cl_g$, e.g. in Error type I, we implicitly assume that neither of them is Nan. We define $Ac_g(i,j)$ (resp. $Ac_o(i,j)$) to be 1, $-1$ or 0 depending on whether the clusters to which $i$ and $j$ belong to are, on the ground truth (resp. output) tree, in a parent-descendant relation, descendant-parent relation, or otherwise. If either $i$ or $j$ is not in any cluster in the output tree, then we set $Ac_o(i,j)$ to Nan. When comparing $Ac_o$ to $Ac_g$, e.g. in Error type II and III, we implicitly assume that neither of them is Nan. The mathematical definitions of the four error types are as follows:

**Err. Type I**= $|S_1|/\binom{q}{2}$, where $S_1=\{(i,j):(Cl_g(i,j)\neq Cl_o(i,j)\}$;
**Err. Type II**= $|S_2|/\binom{q}{2}$, where $S_2=\{(i,j):Ac_g(i,j)\neq Ac_o(i,j)\}$;
**Err. Type III**= $|S_3|/\binom{q}{2}$, where $S_3=\{(i,j):|Ac_g(i,j)|+|Ac_o(i,j)|=1\}$;
**Err. Type IV**= $|S_4|/\binom{q}{2}$, where $S_4=\{(i,j):Cl_o(i,j)$ is Nan$\}$;

We begin by comparing the performance of different phylogenetic tree inference algorithms using Error Type II, which measures the fraction of correctly identified ancestral relations.

Figure 2 shows the distribution of Type II errors over the data sets that the different tools were able to process (c.f Table 1). The mean error and the standard deviation of mean error (SD) are also shown. The size of the output tree is automatically determined by each tool, for EXACT we used (14). For each tool, we are using as prediction the tree that the tool deems to best explain the data, which we then compare with the ground truth tree. For EXACT this is the tree with the lowest $\mathcal{C}$, for the tree size that optimizes (14). EXACT and PhyloWGS perform visibly better than the other tools. EXACT and PhyloWGS perform almost indistinguishably, to be precise, $0.146$ vs. $0.141$ (based on Error Type II). The reason why EXACT performs slightly worse is because, at lower sequencing coverage (1/3 of the input files), the simple kmeans that we use to cluster mutations (and which is not the focus of this paper) is imprecise, and gives our core algorithm, (the main focus of this paper), a disadvantageous starting point. In fact, if we only focus on high and medium sequencing coverage, i.e, $1000\times$ and $100\times$, the mean Type II errors $\pm$ SD of EXACT and PhyloWGS are $\mathbf{0.13 \pm 0.01}$ and $\mathbf{0.14 \pm 0.01}$, respectively.

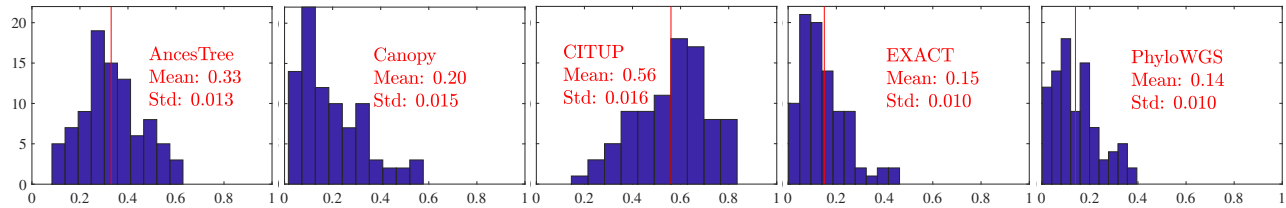

**Fig. 2.** Distribution over all of the 90 inputs files of the Type II errors for the tree output by different algorithms. $x$-axis: error. $y$-axis: count. The vertical red line shows the mean.

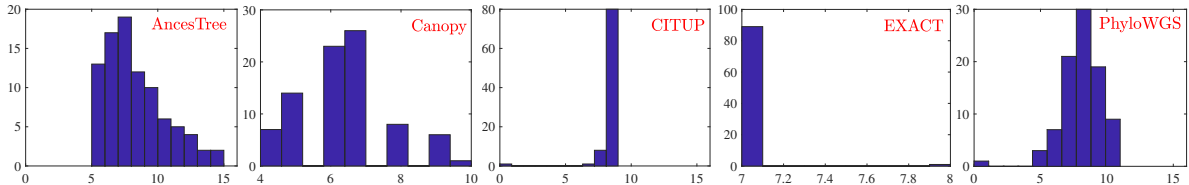

**Fig. 3.** Tree size automatically output by each algorithm for each of the 90 input files.

Table 2 shows the mean error and the standard deviation of mean error for different algorithms and all error types, over all 90 input files. EXACT and PhyloWGS perform similarly, and equal or better than all other tools (except for Error Type I, in which CITUP does slightly better than EXACT). AncesTree is the only algorithm with nonzero Type IV error. On average, half or more of the mutations are missing in the output of AncesTree, while all the mutations are always present in the output trees in all of the other algorithms. Canopy clusters mutations very inaccurately. For the same reasons as discussed in the previous paragraph, if we only focus on high/medium sequencing coverage files, the mean Type III errors of EXACT and PhyloWGS are $\mathbf{0.098 \pm 0.013}$ and $\mathbf{0.102 \pm 0.011}$, respectively.

| Tool | Err. Type I | Err. Type II | Err. Type III | Err. Type IV |
|---|---|---|---|---|
| **AncesTree** | 10.61±0.53 | 33.34±1.34 | 24.72±1.37 | 56.58±0.27 |
| **Canopy** | 18.84±1.38 | 19.95±1.49 | 23.91±2.09 | 0±0 |
| **CITUP** | 7.25±0.43 | 56.36±1.64 | 41.45±1.31 | 0±0 |
| **EXACT** | 8.97±0.62 | 14.64±1.02 | 11.48±1.05 | 0±0 |
| **PhyloWGS** | 6.05±0.49 | 14.08±1.00 | 10.58±0.89 | 0±0 |

Table 2. $10^2 \times$ Mean error $\pm$ SD of the mean, over all 90 input files, for tree output by different algorithms (size determined automatically).

Note that in Table 2, the size of the output tree of EXACT is chosen according to the BIC criterion. To see how different tree sizes from EXACT approximate the ground truth, in Table 3 we show the mean error, $\pm$ the SD of the mean error, for different error types, when, for each tree size, EXACT outputs the tree with the smallest $\mathcal{C}$. Errors do not seem to change much with tree size, the one changing the most being the error of Type II.

Recall that the ground truth tree size does not vary from one input file to the next, it is always 10. In Figure 3, we show the distribution of tree sizes output by the different tools on all input files. EXACT and CITUP show very little variance compared to the other tools.

In Table 4, we provide the distribution of tree sizes output by EXACT that minimize different error types. In other words, for each input file, we generate four output trees of sizes $q \in \{7, 8, 9, 10\}$, each minimizing $\mathcal{C}$ among all the trees of its size. Then, among these four trees, we pick the one with the lowest error for each type of error.

By comparing the histogram for EXACT in Figure 3 to Table 4, we see that the scaling $D$ that we use for EXACT, which affects the BIC criterion and the tree size output, was not overfit to favour EXACT. Yet, EXACT still performs very well. To illustrate EXACT's robustness to a roughly tuned BIC criterion, we recomputed all four error types, as well as the distribution of output tree sizes, when we remove the second term in (14), i.e., we do not penalize large models at all. The mean errors that we get (for Type I to IV) are now 0.079, 0.1481, 0.1071, and 0. The distribution of tree sizes is (61, 22, 4, 3) for trees of sizes (7, 8, 9, 10), respectively.

| Tree size | Err. Type I | Err. Type II | Err. Type III | Err. Type IV |
|---|---|---|---|---|
| **7** | 9.01±0.62 | 14.96±1.00 | 11.84±1.07 | 0±0 |
| **8** | 7.94±0.44 | 15.27±1.01 | 10.93±1.05 | 0±0 |
| **9** | 8.34±0.43 | 16.09±1.01 | 10.68±0.96 | 0±0 |
| **10** | 9.11±0.41 | 16.99±0.82 | 10.60±0.88 | 0±0 |

Table 3. $10^2 \times$ Mean errors $\pm$ SD of the mean, over all 90 input files, for tree with smallest $\mathcal{C}$ when output tree size is predetermined.

| Tree size | Err. Type I | Err. Type II | Err. Type III |
|---|---|---|---|
| **7** | 23 | 33 | 28 |
| **8** | 34 | 29 | 26 |
| **9** | 17 | 17 | 20 |
| **10** | 16 | 11 | 16 |

Table 4. Distribution of sizes of trees with smallest $\mathcal{C}$ that, for each file, have the smallest error of types I-III. Type IV error is always 0 for all tree sizes.

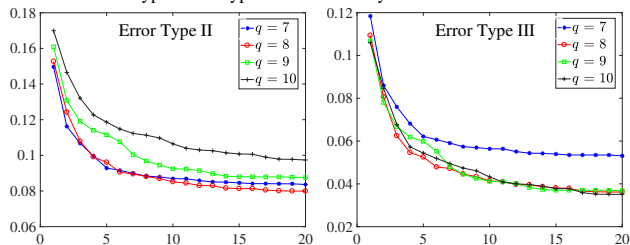

**Fig. 4.** Smallest Type II error ($y$-axis) in the top $k$ ($x$-axis) trees with smallest $\mathcal{C}$ in

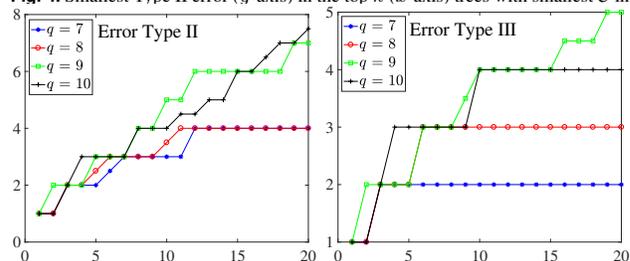

**Fig. 5.** Median ($y$-axis), across all of the 90 synthetic input files, of the rank (in terms of $\mathcal{C}$) of the tree with smallest error (of Types II and III) among the top $k$ ($x$-axis) trees output by EXACT, for different tree sizes

One big advantage of EXACT over competing tools is its ability to accurately output the top $k$ trees with the lowest $\mathcal{C}$ for each tree size. By considering the smallest error among the top $k > 1$ trees, the performance of EXACT surpasses that of PhyloWGS. This is illustrated in Figure 4, which shows the smallest error among the top $k > 1$ trees for Error Types II and III, for different (predetermined) tree sizes, and different values of $k$. In Figure 4, we purposefully do not show curves for Error Types I and IV. Error Type IV is always 0 since the trees output by EXACT never miss any mutations. Error Type I remains constant as we increase $k$, since it only depends on the pre-clustering done by EXACT. The Type I errors for tree sizes of 7, 8, 9, and 10 are 0.079, 0.083, 0.090 and 0.091.

Furthermore, we observe that the best tree among the top $k$ trees tends to have a $\mathcal{C}$ that ranks very close to the minimum $\mathcal{C}$. This is illustrated in Figure 5, which shows, for example, that when EXACT returns the top $k = 10$ trees for size $q = 7$, for more than half of the input files, the tree that minimizes the Type II error is among the top 4 trees with smallest $\mathcal{C}$.

We finish this section by comparing the run-times of different algorithms. Table 5 shows the run-time of EXACT for different tree sizes and various $k$'s. As mentioned earlier, choosing the top $k$ trees only corresponds to sorting the best trees according to their $\mathcal{C}$ values and selecting the top $k$ trees. Therefore, changing $k$ does not affect the overall run-time of EXACT much.

| Top-$k$ \ Tree size | 7 | 8 | 9 | 10 |
|---|---|---|---|---|
| 1 | 1.1 | 12.7 | 303.9 | 8587.1 |
| 21 | 1.1 | 12.9 | 305.9 | 8663.4 |
| 46 | 1.0 | 12.5 | 311.9 | 8621.8 |

Table 5. Run-time (in seconds) when keeping track of the top-$k$ trees with smallest $\mathcal{C}$ in EXACT.

Finally, Table 6 shows the average run-time of all algorithms over all input files. These run-times are obtained when EXACT tests all trees up to size 10 (size 11 if we include the null mutation root node). The average time to solve (7) and compute the cost of each tree is about $3.6\mu$ seconds. Note that we can get the distribution in Fig. 2, or the results of Table 2, within just **12.7** seconds, if EXACT does not test trees of sizes 9 or 10. Regardless, EXACT's run-time is very reasonable, especially considering that it provides more complete and accurate information than the other tools. Its run-time can be further reduced if we use a more recent GPU, multiple GPUs, or more computer cores. Such resource exploitation is not possible with other tools.

| EXACT | AncesTree | Canopy | CITUP | PhyloWGS |
|---|---|---|---|---|
| 8905 | 1.07 | 2642 | 17322 | 2963 |

Table 6. Average run-time (in seconds) of different algorithms.

### 4.2 Results on real data

These data consist of mutation frequencies obtained from patients with (i) chronic lymphocytic leukemia (CLL) via whole genome sequencing (Schuh *et al.*, 2012) (6 files), (ii) lung adenocarcinoma (Zhang *et al.*, 2014) via multiregion sequencing (22 files), and (iii) renal cell carcinoma tumors (Gerlinger *et al.*, 2014) via multiregion sequencing (8 files). Samples in the lung and renal tumors are obtained from distinct positions in the organ, while samples in the CLL tumors are obtained at distinct time points.

We first test the accuracy of the different tools by comparing their output against an expert-built tree (Deshwar *et al.*, 2015) (taken as ground truth) for the *CLL077deep* data set of Schuh *et al.* (2012). The Type II errors (ancestral relations errors) for AncesTree, Canopy, CITUP, EXACT, and PhyloWGS are, respectively, 0.43, 0.24, 0.20, 0.13, 0.08. It is important to look more closely at the two top contenders, EXACT and PhyloWGS. In Fig. 6, we see that both tools recover trees with similar topology, but EXACT explains evolution in a more fine-grained manner by splitting the expert's clusters into several smaller clusters. E.g., the root of both the expert's and PhyloWGS's tree is {SA, BC, NA, GP, SL}, which EXACT splits into {SA}, {BC, NA}, and {GP, SL}. This affects the Type II error because the expert's tree is less granular than EXACT's tree. We can, however, control the granularity in EXACT's output by not using the single most likely tree, but the 100 most likely trees (according the model (2)-(5)). In the bottom-right panel of Fig. 6, we use these trees to estimate the marginal probability of each pair of mutations being in an ancestral relation. Using these probabilities, and clustering in directed graphs (Pentney and Meila, 2005) plus kmeans ($k = 4$), we exactly recover the expert's clusters and topology. Importantly, the recovered $F$ for different samples/nodes is much closer to the expert's values in EXACT than in PhyloWGS, see the top-right panel in Fig. 6.

We also compare the output of different tools against each other, using the error types introduced in Section 4.1. Table 7 shows the mean error, for the four different error types, and for each pair of tools, for all 36 real data

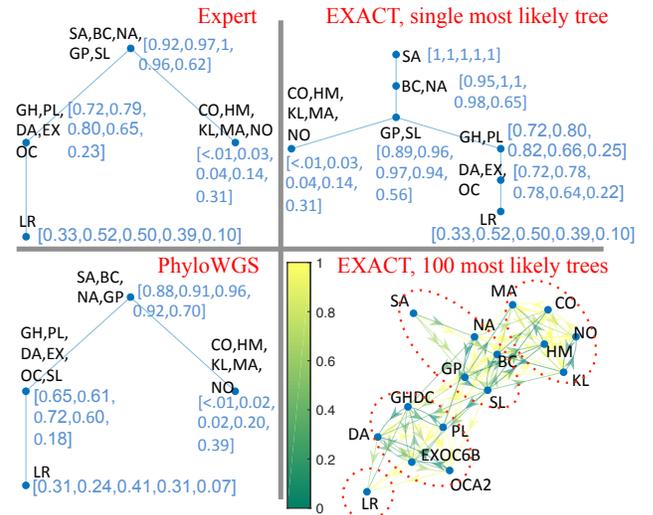

**Fig. 6.** Trees output by EXACT, PhyloWGS, and expert-built tree for the CLL007deep data. The numeric array next to node $i$ is the inferred $F_{i,s}$ across samples $s \in [5]$. The letters next to each node represent a cluster of mutations. All tree roots include SA. In the bottom-right graph, edge-colors are the marginal probability of each two mutations being connected, estimated from the 100 most likely trees output EXACT under assumptions (2)-(5). Red dotted ellipses are clusters of mutations, automatically obtained from the graph via Pentney and Meila (2005) and kmeans ($k = 4$)

sets. Recall that, for each pair of tools, we compare them only on the files that run on both tools (cf. Table 1). In each cell, the top left sub-cell corresponds to Error Type I, the top right sub-cell corresponds to Error Type II, and the bottom left and bottom right sub-cells correspond to Error Type III and Error Type IV, respectively. The standard deviations of the mean for each value in Table 7 range from 0.01 to 0.06.

| | Canopy | | CITUP | | EXACT | | PhyloWGS | |
|---|---|---|---|---|---|---|---|---|
| **AncesTree** | 0.22 | 0.30 | 0.18 | 0.52 | 0.26 | 0.41 | 0.25 | 0.38 |
| | 0.37 | 0.80 | 0.32 | 0.73 | 0.31 | 0.67 | 0.34 | 0.67 |
| **Canopy** | | | 0.09 | 0.48 | 0.11 | 0.21 | 0.10 | 0.21 |
| | | | 0.33 | 0 | 0.18 | 0 | 0.17 | 0 |
| **CITUP** | | | | | 0.11 | 0.51 | 0.12 | 0.55 |
| | | | | | 0.32 | 0 | 0.39 | 0 |
| **EXACT** | | | | | | | 0.15 | 0.29 |
| | | | | | | | 0.27 | 0 |

Table 7. Comparing all tools against each other using the real data sets.

Focusing on Type II errors, i.e., agreement of ancestral relations, we see from Table 7 that AncesTree and CITUP tend not to agree with any other tool, including each other. On the contrary, Canopy, EXACT, and PhyloWGS agree better with each other, Canopy being the tool that agrees most with any other tool. This is consistent with the results in Table 2, which, regarding error Type II, also ranks Canopy, EXACT and PhyloWGS above AncesTree and CITUP. Regarding Type IV errors, i.e., correct clustering of mutations, AncesTree disagrees the most with any other tool. This is again consistent with the results in Table 2, and is because AncesTree leaves many mutations out of its output trees. Regarding Type I, AncesTrees is the tool, whose clustering disagrees the most with that of every other tool, while in Table 2 it is Canopy, which seems to have the worst clustering.

In the Supp. Mat. Section 8 we report additional numerical results for the real data sets.

## 5 Discussion and Future Work

We tested EXACT's performance on previously used benchmark data sets. These data sets are challenging. They have both real and synthetic data, and, not only violate the assumptions (2)-(5), but using them for benchmarking requires that we cluster mutations. Clustering mutations, which we perform using a simple kmeans, is not the focus of our paper, nor is EXACT designed to perform it optimally. Yet, EXACT still performs very well.

EXACT performs better than CITUP, which uses the same pre-clustering procedure and BIC criterion, showing that there is an advantage in doing exact versus approximate inference. EXACT performs better than AncesTree and Canopy, which perform clustering and tree-inference simultaneously. This shows that performing inference carefully, and using a simple clustering method, is better than finding a suboptimal solution to the hard problem of jointly inferring $T$, $M$, and a clustering. Using the most likely tree as a predictor already leads to EXACT having a performance almost as good as PyloWGS's performance on synthetic data (c.f. Fig. 2 and Table 2). Furthermore, accurately sorting all trees by their likelihood, which other tools cannot do, allows EXACT to find a very small ($<$ 3) set of trees among which there is a solution that is much closer to the ground truth than PhyloWGS's prediction on synthetic data (c.f. Fig. 4). This also allows EXACT to output an expert-like tree on real data that is better than PhyloWGS's output (c.f. Fig. 6-(bottom,right)). EXACT reconstructs $M$ values more accurately than PhyloWGS (c.f. Fig. 6-(top,right)).

**Future work:** Other variants of the PPM can benefit from exact inference under noise. One such important case is the PPM for settings where the samples are obtained at different time points, and hence have correlations that can be described by models of cell growth dynamics. In this context, it is convenient to study generalizations of EXACT that solve an extension of (8) where we add $\mathcal{R}(M)$ to the objective, and where $\mathcal{R}(M)$ penalizes unlikely growth dynamics. A good starting point is $\mathcal{R}(M) = C \sum_{s=1}^{p-1} \|M_{.,s} - M_{.,s+1}\|_1$, where $C$ is some constant, and $\|\cdot\|_1$ is the element wise 1-norm. We believe that adding such $\mathcal{R}$ makes the model still exactly solvable in a finite number of steps. These extensions, in addition to potentially leading to less inference bias when the cell population is sampled in time, also have the benefit of reducing variance. Indeed, it can be readily checked that, e.g. under a noiseless setting, the PPM can predict multiple equally good trees, which can be easily distinguished if we look at which of the predicted growth pattern of mutants is more likely or of viable fitness.

Our tool can be applied to infer trees on many different settings. Its good performance on the synthetic data shows that, as long as the infinite-sites assumption approximately holds, EXACT will perform well. We also tested EXACT on 36 cancer data sets previously studied using other tools (Schuh *et al.*, 2012; Zhang *et al.*, 2014; Gerlinger *et al.*, 2014). In the future, we plan to test EXACT on more types of real data, in particular, on VAFs time series data sets obtained from the evolution of antimicrobial resistance. In this context, extending EXACT to consider the addition of $\mathcal{R}$ to (8), as discussed above, might be very useful.

## Funding

This work was partially funded by the following grants: NIH/1U01AI124302, NSF/IIS-1741129, and a NVIDIA hardware grant.

# Supplementary material for "Exact inference under the perfect phylogeny model"

## 6 Table of different tools based on the PPM

In Table 8 we list, to the best of our knowledge, all of the tools that are of the same type as EXACT (i.e. PPM-based accepting as input VAFs) and that are (i) state-of-the-art, and (ii) actively maintained, namely, PhyloSub (Jiao *et al.*, 2014), AncesTree (El-Kebir *et al.*, 2015b), CITUP (Malikic *et al.*, 2015), PhyloWGS (Deshwar *et al.*, 2015), Canopy (Jiang *et al.*, 2016), SPRUCE (El-Kebir *et al.*, 2016), rec-BTP (Hajirasouliha *et al.*, 2014), PASTRI (Satas and Raphael, 2017) and LICHeE (Popic *et al.*, 2015). We also list other tools that are PPM-based but accept as input information other than VAFs. Note that most of these tools, find important applications even while running only on modest-size trees, as can be seen from the column *Tree size* in Table 8.

## 7 Other objective functions

Although the objective (7) looks simple, it allows us to extend our tool to other objectives.

Suppose that we want to define $\mathcal{C}(U; \hat{F}_O)$ as in eq. (15), where $f_{\hat{F}_O}(F)$ is now a generic convex function of $F$, and, for simplicity, $O = [q] \times [p]$.

$$\min_{M, F \in \mathbb{R}^{q \times p}} f_{\hat{F}_O}(F) \text{ s.t. } F = UM, M \geq 0, M^\top \mathbf{1} = \mathbf{1}. \quad (15)$$

We can use our main algorithm as a building block to implement a proximal algorithm, such as the Alternating Direction Method of Multipliers (ADMM), to solve (15). ADMM is guaranteed to converge for convex optimization problems (Boyd *et al.*, 2011), and is a good heuristic to solve complex non-convex problems (Hao *et al.*, 2016; Mathy *et al.*, 2015; Bento *et al.*, 2015; Zoran *et al.*, 2014; Bento *et al.*, 2013). It can achieve a convergence rate equal to the rate of the fastest first-order method for a large class of objective functions (França and Bento, 2016), and quite often it gets close to the optimum in a few steps.

To use ADMM to solve (15), in each iteration we need to solve the following two problems, for an arbitrary $N \in \mathbb{R}^{q \times p}$ and some value $\rho > 0$:

$$\min_{F \in \mathbb{R}^{q \times p}} \frac{\rho}{2} \|F - N\|^2 + f_{\hat{F}_O}(F), \quad (16)$$

$$\min_{M, F \in \mathbb{R}^{q \times p}} \frac{\rho}{2} \|F - N\|^2 \text{ s.t. } F = UM, M \geq 0, M^\top \mathbf{1} = \mathbf{1}. \quad (17)$$

Problem (17) is a special form of problem (7), which our tool allows us to compute very fast. For several interesting functions $f_{\hat{F}_O}$, problem (16) can be solved very fast. E.g., if $f_{\hat{F}_O}(F) = \|\hat{F}_O - F\|_1$, the element-wise 1-norm, then solving (16) amounts to thresholding $N - \hat{F}_O$. Specifically, the minimizer of (16) is $\text{thres}_{1/\rho}(N - \hat{F}_O)$, where $\text{thres}_{1/\rho}()$ acts component-wise, and, for each component $x$, $\text{thres}_{1/\rho}(x) = 0$ if $x \in [-\frac{1}{\rho}, \frac{1}{\rho}]$, it is $x - \frac{1}{\rho}$ if $x > \frac{1}{\rho}$, and it is $x + \frac{1}{\rho}$ if $x < -\frac{1}{\rho}$.

Running a few iterations of ADMM to compute $\mathcal{C}$ for each tree comes at the expense of a slight loss in accuracy, depending on how many ADMM iterations we run, and at the expense of a reasonable extra computational cost, assuming that (16) can be computed as fast as our main algorithm.

## 8 Additional numerical results for real data

The chronic lymphocytic leukemia data sets (CLL) that we used were obtained originally by Schuh *et al.* (2012). Since these authors also provide their own analysis and biological perspectives of these data sets, its is instructive to see how close the reconstructed phylogenies obtained by different tools compare to the original ones.

The work of Schuh *et al.* (2012) reports tumor composition reconstructions for the chronic lymphocytic leukemia data sets CLL003, CLL006 and CLL077, see Figure 8. This allows us to compare how well the different tools recover each of the ratio matrices reported in Fig. 8 using the following procedure. Given matrices $M_1$ and $M_2$, mixing matrices for the same number $p$ of samples but potentially different number of subclones, we start by adding rows of zeros to $M_1$, or $M_2$, representing dummy subclones, such that $M'_1$ and $M'_2$, the padded matrices, have the same number $q$ of rows. Then, we find the permutation matrix $\Pi$ such that when $\Pi$ is applied to the rows of $M'_1$ the distance $\frac{\|\Pi M'_1 - M'_2\|_1}{pq}$ is minimized, where $\|\cdots\|_1$ denotes the entry-wise 1-norm. We report this minimal distance as a performance of how close $M_1$ is to $M_2$ in Table 9. From this table we can see that CITUP recovers mutant mixing ratios that are very similar to those reported in Schuh *et al.* (2012), an observation that had already been reported in Malikic *et al.* (2015). Surprisingly, PhyloWGS, which shows good performance in previous numerical experiments, now seems not to perform that well.

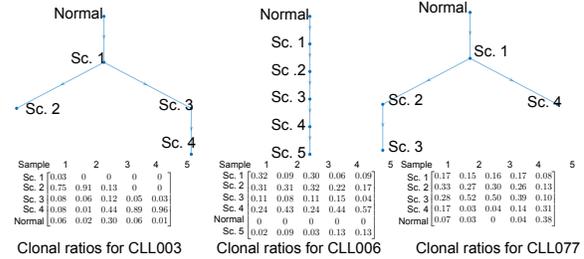

**Fig. 7.** Trees reconstructed by Schuh et al. (2012) for the data sets CLL003, CLL006, and CLL077, and corresponding reconstructed ratios of clonal mixtures. Sc. stands for subclone. For the CLL003 data set, Schuh et al. (2012) also reports clustering of mutations, namely: Sc. 1 = ADAD1, SHROOM1, SF3B1, HERC2, CHTF8, IL11RA; Sc. 2 = ATM, AMTN, APBB2, MTUS1, SPTAN1, PLEKHG5, BPIL2; Sc. 3 = ASXL1, MUSK, SEMA3E; Sc. 4 = CHRNB2, FAT3, NRG3, NPY.

One feature of the $M$ matrix that all tools seem to recover is a switch between two subclones in data set CLL003. In particular, the first sample is always dominated by one subclone while the last sample is always dominated by another subclone. Another feature of the $M$ matrices that we study is the *mixing proportion* (MP), i.e., the fraction of its entries that are non-zero. The authors El-Kebir *et al.* (2015b) report that the $M$ matrices reconstructed by AncesTree have a high MP for the CLL data set but a low MP for the renal and lung tumor data sets. In our own experiments with AncesTree we confirm this observation with values MP of 0.84, 0.23, 0.55 for the CLL, renal, and lung data sets respectively. They claim that these values are reasonable since the CLL deals with liquid tumors, while the other two data sets with solid tumors. However, we observe that other tools only respect this dichotomy with regards to the CLL and the renal tumors data sets. For the lung tumors data sets, AncesTree is the only tool that reports a MP smaller than 0.55. All other tools report a MP larger than 0.73.

For the CLL003 data set, the authors (Schuh *et al.*, 2012) provide their reconstructed clustering of mutations. This allows us to compare the trees output by different algorithms against the predictions in Schuh *et al.* (2012), taken here as ground truth, using the four error types we defined in Section 4. We report these values in Table 10. We see that, EXACT performs quite well, although sometimes slightly worse than Canopy. PhyloWGS also seems to perform relatively well.

| Tool name | Type of input data | Evolution model | Tree size | Model selection | Main algorithm used | Statistics | Resources |
|---|---|---|---|---|---|---|---|
| **EXACT** | VAFs-SNV | PPM | (14, 14) | ML | Exaustive search | Yes | (N, N, N) |
| **AncesTree** | VAFs-SNV | PPM | (17, 10) | ML | ILP | No | (1, N, 0) |
| **BitPhylogeny** | VAFs-SNV *or* DNA Methylation | Restricted-PPM | (30, -) | ML | MCMC | No | (1, 1, 0) |
| **CITUP** | VAFs-(CNV/SNV) | PPM | (8, 6) | ML | Exhaustive Search + QIP | Yes | (1, N, 0) |
| **Canopy** | VAFs-(CNV/SNV) | PPM | (7, 5) | ML | MCMC | No | (1, 1, 0) |
| **LICHeE** | VAFs-SNV | PPM | (110, -) | ML | Constrained search | Yes | (1, 1, 0) |
| **MEDICC** | SNP arrays-CNV | ploidy +/- | (15, 17) | Min. event dist. | Greedy dist. minimization | No | (1, 1, 0) |
| **PASTRI** | read counts-SNV | PPM | (8, 5) | ML | Importance sampling | No | (1, 1, 0) |
| **PhyloWGS** | VAFs-(CNV/SNV) | PPM | (6, 6) | ML | MCMC | Yes | (1, 1, 0) |
| **rec-BTP** | VAFs-SNV | PPM | (9, 9) | ML | Recursive greedy search | No | (1, 1, 0) |
| **SPRUCE** | VAFs-(CNV/SNV) | Multistate-PPM | (21, - ) | Max. Char. Tree | Constrained search | No | (1, 1, 0) |
| **TuMult** | SNP arrays-CNV | ploidy +/- | (12, -) | ML | Iterative greedy merge | No | (1, 1, 0) |
| **FISHtrees** | FISH *or* single cell FISH | ploidy +/- | (54, 160) | ML | MILP | Yes | (1, 1, 0) |
| **oncoNEM** | binary mutation arrays of SNVs | PPM | (22, -) | ML | Constrained search | No | (1, 1, 0) |
| **SCITE** | single cell WGS | PPM | (40, -) | ML | MCMC | No | (1, 1, 0) |

Table 8. Table showing different tools based on the PPM (or variants) to infer trees from data. The last three tools use single-cell sequencing/FISH data, and are of a different type of input data than ours. The other tools use somatic mutation frequency data from multiple samples, just like our method, EXACT.

*Legend for non-self-explanatory notations.* **Tree size** $(x, y)$: largest tree that is processed in the paper introducing each tool for both real data, $x$, and simulated data, $y$; **CNV**: Copy-number variation; **FISH/FISH-SC**: Fluorescence in situ hybridization (SC - from single-cell), with copy-number data; **ploidy +/-**: Model allows copy number increase/decrease of single gene or chromosome; **Model selection**: How the tool chooses which trees to output; **ML**: Method that disambiguates between trees by computing a score, which we can interpret as a maximum likelihood, although these scores are not always introduced via probabilistic arguments; **Statistics**: Does the tool output a distribution/ranking of trees, or just a single best tree (some tools could, with modifications, do it, but current software does not); **Resources** $(x, y, z)$: number of machines, $x$, number of cores in each machine, $y$, and number of GPUs in each machine, $z$, that a tool can use.

|  | AncesTree | EXACT | Canopy | CITUP | PhyloWGS |
|---|---|---|---|---|---|
| **CLL 003** | 0.038 | 0.019 | 0.028 | 0.016 | 0.023 |
| **CLL 006** | 0.01 | 0.009 | 0.008 | 0.003 | 0.058 |
| **CLL 077** | 0.031 | 0.016 | 0.037 | 0.016 | 0.024 |

Table 9. Distribution of minimal distance across different algorithms for CLL 003, CLL 006 and the CLL 077 data sets

|  | AncesTree | Canopy | CITUP | EXACT | PhyloWGS |
|---|---|---|---|---|---|
| **Err. Type I** | 0.2368 | 0.0526 | 0.0316 | 0.0579 | 0.0895 |
| **Err. Type II** | 0.3789 | 0.1789 | 0.7474 | 0.1842 | 0.2158 |
| **Err. Type III** | 0.3737 | 0.1263 | 0.4632 | 0.1263 | 0.1263 |
| **Err. Type IV** | 0.9474 | 0 | 0 | 0 | 0 |

Table 10. Comparison, using four different error types, between the left-most tree in Fig. 8, obtained by Schuh et al. (2012) for the CLL003 data set, and the tree produced by each algorithm on the CLL003 data set.

## 9 Extension of Algorithm 1 to the case when $\mathbf{1}^\top M \leq 1$

To evaluate the cost $\mathcal{C}$ in the case where we choose $\leq$ in constraint (10) in Theorem 3.2, we only need to replace line 12 in Algorithm 1 by

$$t^* = \max\left\{0, t_i - \frac{1 + \mathcal{L}'(t_i)}{\mathcal{L}''(t_i)}\right\}.$$

The justification for this is that, if we write an equivalent version of the dual formulation (11)-(13) for the case when $\mathbf{1}^\top M \leq 1$, we get exactly the same dual formulation but with (13) replaced with $\max\{0, Z_i + N_i\} \leq t \forall i \in \mathcal{V}$, which is exactly the same thing as adding the constraint $t \geq 0$ to the original dual problem.

To see where $\max\{0, Z_i + N_i\} \leq t$ comes from, we can trace first required adjustment to the proof of Lemma 3.3 to equation (33), that now reads $\max_{i\in[q]}\{(U^\top DY)_i, 0\}$, and from which then $\max\{0, Z_i + N_i\} \leq t$ follows.

## 10 Analysis of Algorithm 1

The following lemmas support the explanation and correctness of Algorithm 1 given in Section 3. These lemmas read exactly like the Lemmas 3.2, 3.3, 3.4, 3.7 3.8 and 3.9 stated in Section 3.1 of Jia *et al.* (2018). However, our lemmas pertain a more general problem then Jia *et al.* (2018), i.e., Jia *et al.* (2018) assumes that $D = I$, while we allow $D$ to be a general strictly positive diagonal matrix. The proofs of these lemmas are almost identical to the proofs of the corresponding lemmas in Jia *et al.* (2018), and so we omit them. The only two differences are the following. First, in adapting the proof of Lemma 3.3 (Jia *et al.*, 2018) to the proof of Lemma 10.2, the statement "Since, $\hat{F} = (U^\top)^{-1}N$, is a continuous function of $N$," should read 'Since, $\hat{F} = D^{-2}(U^\top)^{-1}N$, is a continuous function of $N$,". Second, in adapting the proof of Lemma 3.7 (Jia *et al.*, 2018) to the proof of Lemma 10.4, the use of Lemma 3.6 (Jia *et al.*, 2018) should be replace with the use of Lemma 11.2, which we state in our Supp. Mat. Section 11 and which we prove in Supp. Mat. Section 12. Furthermore, the statement "the continuous piecewise quadratic $\mathcal{L}(t) = (1/2)\sum_{i\in\mathcal{V}}(Z^*(t)_i - Z^*(t)_{\bar{i}})^2$." should read "the continuous piecewise quadratic $\mathcal{L}(t) = (1/2)\sum_{i\in\mathcal{V}}D_{ii}^{-2}(Z^*(t)_i - Z^*(t)_{\bar{i}})^2$."

**Lemma 10.1.** *$\mathcal{L}(t)$ is a convex function of $t$ and $N$. Furthermore, $\mathcal{L}(t)$ is continuous in $t$ and $N$, and $\mathcal{L}'(t)$ is non-decreasing with $t$.*

**Lemma 10.2.** *$Z^*(t)$ is continuous as a function of $t$ and $N$. $Z^*(t^*)$ is continuous as a function of $N$.*

Recall that $\mathcal{B}(t) = \{i : Z^*(t)_i = t - N_i\}$, i.e., the set of components of the solution at the boundary of (12). Variables in $\mathcal{B}$ are called *fixed*, and we call other variables *free*. Free (resp. fixed) nodes are nodes corresponding to free (resp. fixed) variables.

**Lemma 10.3.** *$\mathcal{B}(t)$ is piecewise constant in $t$.*

**Lemma 10.4.** *$Z^*(t)$ and $\mathcal{L}'(t)$ are piecewise linear and continuous in $t$. Furthermore, $Z^*(t)$ and $\mathcal{L}'(t)$ change linear segments if and only if $\mathcal{B}(t)$ changes.*

**Lemma 10.5.** *If $t \leq t'$, then $\mathcal{B}(t') \subseteq \mathcal{B}(t)$. In particular, $\mathcal{B}(t)$ changes at most $q$ times with $t$.*

**Lemma 10.6.** *$Z^*(t)$ and $\mathcal{L}'(t)$ have less than $q + 1$ different linear segments.*

With these lemmas in hand, it is not hard to prove Theorem 3.4. Its proof is exactly the same as a combination of the proofs of Theorems 3.10 and 3.11 in Jia *et al.* (2018). The only modifications are the following. First, any invocation to Theorem 3.19 in Jia *et al.* (2018) should be replaced with an invocation to Theorem 11.7. Second, any invocation to Lemmas 3.1, 3.2, 3.7, 3.8, 3.9 in Jia *et al.* (2018) should be replaced with an invocation to Theorem 3.3, and Lemmas 10.1, 10.4, 10.5, 10.6 respectively.

## 11 Computing rates

We now explain how the procedure *ComputeRates* works. This is an extension of the procedure by the same name introduced in Jia *et al.* (2018). Recall that this procedure takes as input the tree $T$ (which does not change as Algorithm 1 runs) and the set $\mathcal{B}(t)$, and it outputs the derivatives $Z'^*(t)$ and $\mathcal{L}''^*(t)$. We recall that $\mathcal{B}(t) = \{r : Z^*(t)_r = t - N_r\}$, is the set of nodes for which (13) is active for the inner problem (12). Variables in $\mathcal{B}$ are called *fixed*, and we call other variables *free*. Free (resp. fixed) nodes are nodes corresponding to free (resp. fixed) variables.

There are two main ideas behind the procedure ComputeRates. The first is to notice that, for a given $t$, problem (12), which depends on the tree $T$, decomposes into a series of smaller independent problems, each depending on a subtree of $T$, and involving a disjoint subset of the optimization variables $\{Z(t)_r\}$. This decomposition only depends on $B(t)$, which does not change in a small interval $(t - \epsilon, t]$, $\epsilon > 0$, which includes the $t$ for which we want to compute the rates (cf. Lemma 10.5 in App. 10). The second is to notice that each of these sub-problems is a simple unconstrained convex quadratic problem, which can be solved in closed form, and from which $Z(s)$, $\mathcal{L}(s)$, $s \in (t - \epsilon, t]$, and their derivatives, can be computed. We now detail these two ideas.

Consider dividing the tree $T = (r, \mathcal{V}, \mathcal{E})$, for which we want to compute the cost $\mathcal{C}$, into subtrees, each with at least one free node, using $\mathcal{B}(t)$ as separation points. See Figure 8 for an illustration. Each $i \in \mathcal{B}(t)$ belongs to at most degree$(i)$ different subtrees, where degree$(i)$ is the degree of node $i$, and each $i \in \mathcal{V}\backslash\mathcal{B}(t)$, i.e., each free node, belongs exactly to one subtree. Let $\{T_w\}_{w \in [k]}$ be the set (of size $k$) of resulting (rooted, labeled) trees. Let $T_w = (r_w, \mathcal{V}_w, \mathcal{E}_w)$, where the root $r_w$ is the closest node in $T_w$ to $r$. We call $\{T_w\}$ the subtrees *induced by* $\mathcal{B}(t)$. We define $\mathcal{B}_w(t) = \mathcal{B}(t) \cap \mathcal{V}_w$, and, when it does not create ambiguity, we drop the index $t$ in $\mathcal{B}_w(t)$. Note that different $\mathcal{B}_w(t)$'s might have elements in common. Also note that, by construction, if $i \in \mathcal{B}_w$, then $i$ must be a leaf of $T_w$, or the root of $T_w$.

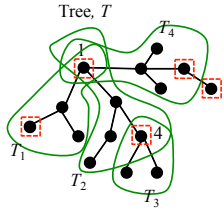

**Fig. 8.** Four subtrees of $T$, with root 1, induced by $\mathcal{B}(t)$, the fixed nodes, represented by the red squares. The root of $T_1$, $T_2$ and $T_4$ is node 1. The root of $T_3$ is node 4. All subtrees must have nodes associated to free variables (free nodes). Within each subtree, any fixed node must be the root or a leaf.

**Definition 11.1.** *The $(T_w, \mathcal{B}_w)$-problem is the optimization problem over $|\mathcal{V}_w\backslash\mathcal{B}(t)|$ variables*

$$\min_{\{Z_j : j \in \mathcal{V}_w \backslash \mathcal{B}(t)\}} \frac{1}{2} \sum_{j \in \mathcal{V}_w} D_{jj}^{-2} (Z_j - Z_{\bar{j}})^2, \qquad (18)$$

*where $\bar{j}$ is the parent of $j$ in $T_w$, $Z_{\bar{j}} = 0$ if $j = r_w$, and $Z_j = Z^*(t)_j = t - N_j$ if $j \in \mathcal{B}_w(t)$.*

**Lemma 11.2.** *Problem (12) decomposes into $k$ independent problems. In particular, the minimizers $\{Z^*(t)_j : j \in \mathcal{V}_w \backslash \mathcal{B}(t)\}$ are determined as the solution of the $(T_w, \mathcal{B}_w)$-problem. If $j \in \mathcal{V}_w$, then $Z^*(t)_j = c_1 t + c_2$, where $c_1$ and $c_2$ depend on $j$ but not on $t$, and $0 \leq c_1 \leq 1$.*

The proof of Lemma 11.2 is in Section 12 in the Supp. Mat..

We note that if $j \in \mathcal{B}(t)$, then, by definition, $Z'^*(t)_j = 1$, therefore, to compute $Z'(t)$, we only need focus on components $j \in \mathcal{V}\backslash\mathcal{B}(t)$. Lemma 11.2 implies that we can find $Z'^*(t)_j$ by solving the $(T_w = (r_w, \mathcal{V}_w, \mathcal{E}_w), \mathcal{B}_w)$-problem as a function of $t$, where $w$ is such that $j \in \mathcal{V}_w \backslash \mathcal{B}_w$. Furthermore, as the next lemma shows, once we have all $\{Z'^*(t)_j\}_{j \in \mathcal{V}}$, we can easily compute $\mathcal{L}''^*(t)$.

**Lemma 11.3.** *$\mathcal{L}''(t)$ can be computed from $Z'^*(t)$ in $\mathcal{O}(q)$ steps and with $\mathcal{O}(1)$ memory as*

$$\mathcal{L}''(t) = \sum_{j \in \mathcal{V}} D_{jj}^{-2} (Z'^*(t)_j - Z'^*(t)_{\bar{j}})^2, \qquad (19)$$

*where $\bar{j}$ is the closest ancestor to $j$ in $T$.* The proof of Lemma 11.3 amounts to taking two derivatives of the objective in (12), and is a very simple modification of the proof of Lemma 3.12 of Jia *et al.* (2018), which is in Supp. Mat. E, pag. 16, in Jia *et al.* (2018).

We will solve a $(T_w, \mathcal{B}_w)$-problem by recursively solving a slightly more general problem, which we define next. To simplify notation, in the rest of this section, we refer to $T_w$ and $\mathcal{B}_w$ as $T$ and $\mathcal{B}$. Recall that, by the definition of $T$ (i.e., $= T_w$) and of $\mathcal{B}$ (i.e., $= \mathcal{B}_w$), if $i \in \mathcal{B}$, then $i$ must be a leaf of $T$, or the root of $T$.

**Definition 11.4.** *Consider a rooted tree $T = (r, \mathcal{V}, \mathcal{E})$, a set $\mathcal{B} \subseteq \mathcal{V}$, and variables $\{Z_j : j \in \mathcal{V}\}$ such that, if $j \in \mathcal{B}$, then $Z_j = \alpha_j t + \beta_j$ for some $\alpha$ and $\beta$. We define the $(T, \mathcal{B}, \alpha, \beta, \gamma)$-problem as*

$$\min_{\{Z_j : j \in \mathcal{V}\backslash \mathcal{B}\}} \frac{1}{2} \sum_{j \in \mathcal{V}} \gamma_j (Z_j - Z_{\bar{j}})^2, \qquad (20)$$

*where $\gamma > 0$, $\bar{j}$ is the closest ancestor to $j$ in $T$, and $Z_{\bar{j}} = 0$ if $j = r$.*

We refer to the solution of the $(T, \mathcal{B}, \alpha, \beta, \gamma)$-problem as $\{Z_j^* : j \in \mathcal{V}\backslash\mathcal{B}\}$, which uniquely minimizes (20). Note that (20) is unconstrained and its solution, $Z^*$, is a linear function of $t$. Furthermore, the $(T_w, \mathcal{B}_w)$-problem is the same as the $(T_w, \mathcal{B}_w, \text{diag}(D^{-2}), -N, \mathbf{1})$-problem. We are only interested in the rate of change $Z'^*(t)$. It turns out that we actually only need to solve the $(T_w, \mathcal{B}_w, \text{diag}(D^{-2}), 0, \mathbf{1})$-problem, i.e., we can ignore $N$ when computing rates. This is justified by the following lemma, whose proof is a trivial extension of the proof of Lemma 3.18 in Jia *et al.* (2018).

**Lemma 11.5.** *Let $Z^*(t)$ be the solution of the $(T, \mathcal{B})$-problem, and let $\tilde{Z}^*(t)$ be the solution of the $(T, \mathcal{B}, \text{diag}(D^{-2}), 0, \mathbf{1})$-problem. Then, $Z^*(t) = c_1 t + c_2$, for some $c_2$ and, where $c_1$ is such that $\tilde{Z}^*(t) = c_1 t$.*

The recursive procedure to solve a generic $(T, \mathcal{B}, \alpha, \beta, \gamma)$-problem is given by Algorithm 2 in Jia *et al.* (2018), which they call *ComputeRatesRec*. Note that although this procedure is intrinsically recursive, we have implemented it in C in a non-recursive form, which brings us additional execution speed. As proved in Lemma 3.17 of Jia *et al.* (2018), ComputeRatesRec solves such a problem in $\mathcal{O}(q)$ steps and using $\mathcal{O}(q)$ memory, where $q$ is the number of nodes in $T$. ComputeRatesRec assumes that the leafs of $T$ are all fixed, and since $T$ might not satisfy this property, they introduce the following lemma.

Lemma 11.6 (Lemma 3.14 Jia *et al.* (2018), pag. 5). *Consider the solution $Z^*$ of the $(T, \mathcal{B}, \alpha, \beta, \gamma)$-problem. Let $j \in \mathcal{V}\backslash\mathcal{B}$ be a leaf. Then $Z_j^* = Z_{\tilde{j}}^*$. Furthermore, consider the $(\tilde{T}, \mathcal{B}, \alpha, \beta, \gamma)$-problem, where $\tilde{T} = (\tilde{r}, \tilde{\mathcal{V}}, \tilde{\mathcal{E}})$ is equal to $T$ with node $j$ pruned, and let its solution be $\tilde{Z}^*$. We have that $Z_i^* = \tilde{Z}_i^*$, for all $i \in \tilde{\mathcal{V}}$.*

Finally, we can state the full procedure to ComputeRates, which we use in Algorithm 1.

---

**Algorithm 2** ComputeRates (input: $T$ and $\mathcal{B}(t_i)$ output: $Z'^*(t_i)$ and $\mathcal{L}''(t_i)$)

---
1: $Z'^*(t_i)_j = 1$ for all $j \in \mathcal{B}(t_i)$
2: **for** each $(T_w, \mathcal{B}_w)$-problem induced by $\mathcal{B}(t_i)$ **do**
3:    Set $\tilde{T}_w$ to be $T_w$ pruned of all leaf nodes in $\mathcal{B}_w$, by repeatedly evoking Lemma 11.6
4:    ComputeRatesRec($\tilde{T}_w, j, \mathcal{B}_w, \text{diag}(D^{-2}), \mathbf{0}, \mathbf{1}, \tilde{Z'}^*$) by using Algorithm 2 in Jia *et al.* (2018)
5:    $Z'^*(t_i)_j = \tilde{Z'}^*_j$ for all $j \in V_w\backslash\mathcal{B}$
6: **end for**
7: Compute $\mathcal{L}''(t_i)$ from $Z'^*(t_i)$ using Lemma 11.3
8: **return** $Z'^*(t_i)$ and $\mathcal{L}''(t_i)$

---

Theorem 11.7. *Algorithm 2 correctly computes $Z'^*(t_i)$ and $\mathcal{L}''(t_i)$ in $\mathcal{O}(q)$ steps, and uses $\mathcal{O}(q)$ memory, where $q$ is the number of nodes of the input tree.*

The correctness of Algorithm 2 follows from the correctness of the procedure CompueRatesRec, and the correctness of Lemmas 11.2, 11.3 and 11.6. The proof of correctness for Theorem 11.7 is exactly the proof of correctness for Theorem 3.19 in Jia *et al.* (2018). The proof of the run-time/memory bounds in Theorem 11.7 is exactly the same as the proof used in proving the same bounds for the ComputeRates algorithm of Jia *et al.* (2018), given by Theorem 3.19 in Jia *et al.* (2018).

## 12 Proofs

Proof of Lemma 3.1. We assume that the tree has $q$ nodes. The matrix $T$ is such that $T_{v,v'} = 1$ if and only if $v$ is the closet ancestor of $v'$. Because of this, the $v$th column of $T^k$ has a one in row $v'$ if and only if $v'$ is an ancestor $v$ separated by $k$ generations. Thus, the $v$th column of $I + T + T^2 + \cdots + T^{q-1}$, contains a one in all the rows $v'$ such that $v'$ is an ancestor of $v$, or if $v = v'$. But this is the definition of the matrix $U$ associated to the tree $T$. Since no two mutants can be separated by more than $q - 1$ generations, $T^k = 0$ for all $k \geq q$. It follows that

$$U = I + T + T^2 + \cdots + T^{q-1} = \sum_{i=0}^{\infty} T^i = (I - T)^{-1}.$$

Proof of Theorem 3.2. Since $F_H$ does not appear in problem (7), the constraint $F = UM$ can be replaced by $F_O = (UM)_O$. Further replacing the definition of $\tilde{M}_O$ into this new problem, and defining $\tilde{U} = (U_{O,O})^{-1}U_{O,H}$, leads to

$$\mathcal{C}(U; \hat{F}_O) = \min_{M, F \in \mathbb{R}^q} \|D(\hat{F}_O - F_O)\|^2 \qquad (21)$$

subject to

$$F_O = (UM)_O = U_{O,O}M_O + U_{O,H}M_H = U_{O,O}\tilde{M}_O, \quad (22)$$

$$M_H \geq 0, \qquad (23)$$

$$\tilde{M}_O \geq \tilde{U}M_H, \qquad (24)$$

$$\tilde{M}_O^\top \mathbf{1} = \mathbf{1} - (\mathbf{1}^\top - \mathbf{1}^\top \tilde{U})M_H. \qquad (25)$$

To prove Theorem 3.2 it is enough to prove that (i) $\tilde{U}$ is a binary matrix with at most one 1 in each column, and that (ii) $\tilde{U}$ has columns with no 1's if and only if $O$ does not include the root of the tree.

Indeed, if (i) is true, then each component of $\tilde{U}M_H$ involves a sum of different components of $M_H$, and these components do not show up in $(\mathbf{1}^\top - \mathbf{1}^\top \tilde{U})M_H$. Therefore, we can replace (23)-(24) by $\tilde{M}_O \geq 0$. If (ii) is true, then, $(\mathbf{1}^\top - \mathbf{1}^\top \tilde{U})M_H = 0$ if $O$ includes the root of the tree, and, otherwise, it is equal to a sum of components of $M_H$ that do not show up in $\tilde{U}M_H$. Therefore, we can replace (25) by $\tilde{M}_O^\top \mathbf{1} = 1$ or $\tilde{M}_O^\top \mathbf{1} \leq 1$, respectively.

To prove (i) and (ii) notice that $U_{O,O}$ is the ancestry matrix of a tree $T'$ over the nodes $O$ that can be obtained from the original tree $T$ by jumping over all the nodes not in $O$. Notice also that $U_{O,H}$ is a binary matrix that, for column $j \in H$, has a 1 in all rows $i \in O$ for which $i$ is an ancestor of $j$ in $T$.

By Lemma 3.1 we have $(U_{O,O})^{-1} = I - T'$, where, overloading notation, $T'$ represents a binary matrix whose rows and columns we can index by the elements of $O$, and whose row $i \in O$ has a 1 in for each column $j \in O$ such that $j$ is a child of $i$ in the tree $T'$. Therefore, the column vector $T'(U_{O,H})_{.,j}$ has a 1 for each row $i \in O$ for which $i$ is the father of an ancestor of $j$ in $O$. Hence, the column vector $(I - T')(U_{O,H})_{.,j}$ has a 1 in the row $i \in O$ corresponding to the closest ancestor in $O$ of $j \in H$. Therefore, column $j$ of $\tilde{U}$ has at most one 1, and it has no ones if and only if $j$ has no ancestors in $O$, which can happen if and only if the root of $T$ is not in $O$.

Proof of Theorem 3.3. Our proof is an extension of the proof of Theorem 3.1 in Jia *et al.* (2018) to the case $D \neq I$.

Before we start, we note that we drop the subscript $_O$ in $\hat{F}_O$, and we will also assume that $p = 1$, since when $p > 1$ the problem decomposes into $p$ independent smaller problems. With this in mind, the problem to which we want to prove equivalence is

$$\min_{M \in \mathbb{R}^q} \frac{1}{2}\|D\hat{F} - DUM\|^2 \qquad (26)$$

subject to $M \geq 0, M^\top \mathbf{1} = 1$.

We start by making the changes of variables $\tilde{F} = D\hat{F}$ and $\tilde{M} = DUM$, after which the problem reads

$$\min_{\tilde{M} \in \mathbb{R}^q} \frac{1}{2}\|\tilde{M} - \tilde{F}\|^2 \qquad (27)$$

subject to $U^{-1}D^{-1}\tilde{M} \geq 0, \mathbf{1}^\top U^{-1}D^{-1}\tilde{M} = 1$.

We now define the following indicator function

$$g(\tilde{M}) = \begin{cases} 0, & \text{if } (U^{-1}D^{-1}\tilde{M}) \geq 0 \text{ and } \mathbf{1}^\top U^{-1}D^{-1}\tilde{M} = 1, \\ +\infty, & \text{otherwise,} \end{cases}$$

and once again re-write our problem, this time as computing

$$G(\tilde{F}) = \arg\min_{\tilde{M} \in \mathbb{R}^q} g(\tilde{M}) + \frac{1}{2}\|\tilde{M} - \tilde{F}\|^2. \qquad (28)$$

We now notice that computing (28) is computing the *proximal operator* associated with $g$. In turn, since $g$ is a convex, closed and proper function in the extended reals, we can compute $G(\tilde{F})$ by computing the proximal operator, $G^*(\tilde{F})$, of the Fenchel dual of $g$, denoted by $g^*$, and then using Moreau's decomposition (Parikh *et al.*, 2014), namely, $G(\tilde{F}) = \tilde{F} - G^*(\tilde{F})$.

To be concrete, the definition of $G^*(\tilde{F})$ is

$$G^*(\tilde{F}) = \arg\min_{Y \in \mathbb{R}^q} g^*(Y) + \frac{1}{2}\|Y - \tilde{F}\|, \quad (29)$$

where $g^*$ is defined as

$$g^*(Y) = \sup_{\tilde{M} \in \mathbb{R}^q} \{Y^\top \tilde{M} - g(\tilde{M})\} \quad (30)$$

$$= \max_{M \in \mathbb{R}^q} (U^\top DY)^\top M \quad (31)$$

$$\text{subject to } M \geq 0, \mathbf{1}^\top M = 1 \quad (32)$$

$$= \max_{i \in [q]} (U^\top DY)_i. \quad (33)$$

We can thus write $G^*(\tilde{F})$ as

$$G^*(\tilde{F}) = \arg\min_{Y \in \mathbb{R}^q} \max_{i \in [q]} (U^\top DY)_i + \frac{1}{2}\|Y - \tilde{F}\|^2 \quad (34)$$

$$= \arg\min_{Y \in \mathbb{R}^q, t \in \mathbb{R}} t + \frac{1}{2}\|Y - \tilde{F}\|^2 \quad (35)$$

$$\text{subject to } U^\top DY \leq t.$$

Making the change of variable $Z = U^\top D(Y - \tilde{F})$, we can re-write $G^*(\tilde{F})$, and therefore $G(\tilde{F})$, as

$$G^*(\tilde{F}) = \tilde{F} + D^{-1}(U^{-1})^\top Z^* \text{ and } G(\tilde{F}) = -D^{-1}(U^{-1})^\top Z^*, \quad (36)$$

where

$$(Z^*, t^*) = \arg\min_{Z \in \mathbb{R}^q, t \in \mathbb{R}} t + \frac{1}{2}\|D^{-1}(U^{-1})^\top Z\|^2 \quad (37)$$

$$\text{subject to } Z + U^\top D^2 \hat{F} \leq t.$$

To finish the proof we make the following observations.

First, notice that $((U^{-1})^\top Z)_i = ((I - T^\top)Z)_i = Z_i - Z_{\bar{i}}$, where $\bar{i}$ is the father of $i$ in the tree $T$. Second, notice that $(U^T D^2 \hat{F})_i = N_i = \sum_{j \in \Delta i} D_{ii}^2 \hat{F}_j$, where $\Delta i$ is the set of all ancestors of $i$ in $T$, plus $i$. Third, recall that $G(\tilde{F})$ is an optimal value of $-DUM^*$, where $M^*$ is an optimal value for the distribution of mutants. Therefore, $M^* = -U^{-1}D^{-2}(U^{-1})^\top Z^* = -(I - T)D^{-2}(I - T^\top)Z^* = -D^{-2}(I - T^\top)Z^* + TD^{-2}(I - T^\top)Z^*$, from which it follows that $M_i^* = D_{ii}^{-2}(-Z_i^* + Z_{\bar{i}}^*) + \sum_{r \in \partial i} D_{rr}^{-2}(Z_r^* - Z_{\bar{i}}^*)$, where $\partial i$ is the set of children of $i$ in $T$. Fourth, $(F^*)_i = (UM^*)_i = D_{ii}^{-1}(-Z_i^* + Z_{\bar{i}}^*)$.

To see that $M^*$ and $F^*$ are unique, notice that problem (7) is a projection onto a convex set polytope, which always has a unique minimizer. Moureau's decomposition implies that $G^*(\tilde{F})$ is unique, hence the minimizer $Y^*$ of (29) is unique. Thus, $Z^* = U^\top D(Y^* - \hat{F})$ and $t^* = g^*(Y^*)$ are also unique.

Proof of Lemma 11.2. First note that, by definition of $\mathcal{B}(t)$, we know the optimal value $Z_i^*$ of all nodes in $i \in \mathcal{B}(t)$. Hence, the unknowns in problem (12)-(13) are the variables in $\mathcal{V}\setminus\mathcal{B}(t)$, which can be partitioned into disjoint sets $\{\mathcal{V}_w\setminus\mathcal{B}(t)\}_{w=1}^k$.

Second notice that for each term in the objective (12) that involves not known variables, there is some subtree $T_w$ that contains both of its variables. It follows that, given $\mathcal{B}(t)$, problem (12) breaks into $k$ independent problems, the $w$th problem having as unknowns only the variables in $\mathcal{V}_w\setminus\mathcal{B}(t)$ and all terms in the objective where either $j$ or $\bar{j}$ are in $\mathcal{V}_w\setminus\mathcal{B}(t)$.

Obviously, if $j \in \mathcal{V}_w \cap \mathcal{B}(t) \triangleq \mathcal{B}_w(t)$, then, by definition, $Z^*(t)_j = c_1 t + c_2$, with $c_1 = 1$. To find the behavior of $Z^*(t)_j$ for $j \in \mathcal{V}_w\setminus\mathcal{B}(t)$, we need to solve (18). To solve (18), notice that the first-order optimally conditions imply that, if $j \in \mathcal{V}_w\setminus\mathcal{B}(t)$, then

$$Z_j = \frac{1}{C}\left(D_{jj}^{-2}Z_{\bar{j}} + \sum_{r \in \Delta j \setminus \{j\}} D_{rr}^{-2}Z_r\right), \quad (38)$$

where, $C = D_{jj}^{-2} + \sum_{r \in \Delta j \setminus \{j\}} D_{rr}^{-2}$, and, we recall, $\Delta j$ denotes the children of node $j$, plus $j$, in $T_w$.

The variables on the right hand side of (38) that are in $\mathcal{B}_w(t)$ are of the form $t + c$, for some constant $t$, therefore, the set of equations (38) for the free variables amounts to solving a linear system of the kind $A(\{Z_j\}_{j \in \mathcal{B}_w(t)}) = t + b$, for some matrix $A$ and vector $b$.

It follows that $Z_j = c_1 t + c_2$, for some $c_1$ and $c_2$ that depend on $T$, $N$ and $\mathcal{B}$. If we solve for $Z_j$ by recursively applying (38), which only has positive coefficients on the right hand side, it is immediate to see that $c_1 \geq 0$.

To see that $c_1 \leq 1$, we study how $Z_j$, defined by (38), depends on $t$ algebraically. To do so, we treat $t$ as a variable. The study of this algebraic dependency in the proof should not be confused with $t$ being fixed in the statement of the theorem.

Define $\rho = \frac{\sum_{r \in \partial j \cap \mathcal{B}_w(t)} D_{rr}^{-2}}{C}$, where $\partial j$ are the neighbors of $j$ in $T_w$ and notice that

$$\max_j\{Z_j\} \leq \rho t + (1 - \rho)\max_j\{Z_j\} + C, \quad (39)$$

in which $C$ is some constant. Recursively applying the above inequality we get

$$\max_j\{Z_j\} \leq t + C', \quad (40)$$

in which $C'$ is some constant. This shows that no $Z_j$ can grow with $t$ faster than $1 \times t$ and hence $c_1 \leq 1$.

## 13 Details on the failure of previous software

The tests in Table 1 were done using a version of AncesTree, PhyloWGS, and CITUP, downloaded from
`https://github.com/raphael-group/AncesTree`,
`https://github.com/morrislab/phylowgs`,
and `https://github.com/sfu-compbio/citup`,
on June 2019, respectively. For Canopy, the tool was downloaded from
`https://github.com/yuchaojiang/Canopy`
in November 2018.

On the files that are listed as failing: AncesTree ran for more than 5 hours, after which we terminated the program; PhyloWGS produces an empty output file; CITUP terminates with a *segmentation fault*; Canopy terminates with a *subscript out of bounds* error.